%% file: brag0427_arxiv.tex
\documentclass[12pt,preprint]{aastex}
\shorttitle{Characterization of a B stars sample}
\shortauthors{Bragan\c ca {\it et al.}}

\begin{document}

\title{
Projected Rotational Velocities and Stellar Characterization of 350 B Stars in the Nearby Galactic Disk}

\author{G. A. Bragan\c ca\altaffilmark{1}, S. Daflon}
\affil{Observat\'orio Nacional-MCT, Rua Jos\'e Cristino, 77. CEP 20921-400 , Rio de Janeiro-RJ, Brazil}

\and
\author{K. Cunha}
\affil{Observat\'orio Nacional-MCT, Rua Jos\'e Cristino, 77. CEP 20921-400 , Rio de Janeiro-RJ, Brazil}
\affil{National Optical Astronomy Observatory, 	950 North Cherry Avenue , Tucson, AZ 85719, USA}
\affil{University of Arizona, Tucson, AZ, USA}

\and
\author{T. Bensby}
\affil{Lund Observatory, Department of Astronomy and Theoretical Physics, Box 43, SE-22100, Lund, Sweden}

\and
\author{M. S. Oey}
\affil{Astronomy Department, University of Michigan, 830 Dennison Building, Ann Arbor, MI 48109-1042, USA;}

\and
\author{G. Walth}
\affil{Steward Observatory, University of Arizona, 933 N. Cherry Ave, Tucson, AZ 85721, USA }

\altaffiltext{1}{braganca@on.br}

\clearpage

\begin{abstract}
Projected rotational velocities ($v\sin i$) are presented for a sample of 350 early B-type main sequence stars in
the nearby Galactic disk. The stars are located within $\sim 1.5$  kpc from the Sun, and the great majority within 700 pc. The analysis is based on high-resolution spectra obtained with the MIKE
spectrograph on the Magellan Clay 6.5-m telescope at the Las Campanas Observatory in Chile.
Spectral types were estimated based on relative intensities of some key line absorption ratios and comparisons to
synthetic spectra. Effective temperatures were estimated from the reddening-free $Q$ index, and
projected rotational velocities were then determined via interpolation on  a published grid that correlates the synthetic full width at half maximum of the
He {\sc i} lines at $\lambda\lambda4026$, 4388 and 4471 ̊\AA\ with $v\sin i$.  As the sample has been selected solely on the basis of spectral
types it contains an selection of B stars in the field, in clusters, and in OB associations. 
The $v\sin i$ distribution obtained for the entire sample is found to be essentially flat for $v\sin i$ values between 0 -- 150 km s$^{-1}$, with only a modest peak at low projected rotational velocities.
Considering subsamples of stars, there appears to be a gradation
in the $v\sin i$ distribution with the field stars presenting a larger fraction of the slow rotators and the cluster stars distribution showing
an excess of stars with $v\sin i$ between 70 and 130 km s$^{-1}$. Furthermore, for a subsample of potential runaway stars 
we find that the $v\sin i$ distribution resembles the distribution seen in denser environments,  which could suggest that these
runaway stars have been subject to dynamical ejection mechanisms.
\end{abstract}


 \section{Introduction}

O and B type stars, with typical values of projected rotational velocities ($v \sin i$) around  100 km\,s$^{-1}$ and higher,   have the largest average $v \sin i$ values among all main-sequence stars.  Stellar rotation appears to be a fundamental parameter constraining  the formation of these massive stars and  the environments in which they are born,  as well as their subsequent  evolution. For instance, there is observational evidence that stars formed in denser environments tend to rotate faster than those formed in associations  \citep{wolff07} and for O and B stars in the field the proportion of slow rotators seems to be even higher (see \citealt{hg06a} for open clusters and \citealt{daflon07} for the Cep OB2 association). In addition, rotation may modulate the formation of massive field stars. \cite{OL11} cite this trend, together with additional empirical evidence based on the stellar clustering law, IMF, and direct observations, as evidence that significant numbers of field massive stars form {\it in situ}, i.e., they
were not born in clusters. Also, rotation might help in understanding the origin of runaway stars. $V \sin i$ distributions of runaway stars have not been much studied in the literature. \cite{martin06} studied the $v\sin i$ distribution of high latitude OB runaway stars and noted the lack of slow rotators compared to a field sample. This was interpreted in that study as evidence that those runaway stars might have been ejected from OB associations.

The study of $v \sin i$ distributions of samples of OB stars born in different environments, such as clusters, OB associations or the general Galactic field, and selected without bias concerning cluster membership, can be used to probe the interplay between star formation and  stellar rotation. 
In this paper we analyse such a sample; we present the spectroscopic observations  and a first characterization of 
a sample of 350 OB stars located within $\sim$ 2 kpc from the Sun. 
The goal of this study is to define the stars in terms of their effective temperatures, along with their projected rotational velocities, with the emphasis on the $v\sin i$ distributions from stars in different environments. 
These stars will be analysed in terms of their chemical composition in a future study.
This paper is divided as follow: Sect. \ref{observation} describes the observations and sample selection; Sect. \ref{binarity} selects from the observed sample the binary or multiple stars; Sect. \ref{spectral} discusses the derived effective temperatures and spectral classification for the sample. Finally, projected rotational velocities are derived in Sect. \ref{velocity}. In Sect. \ref{discussion} we discuss the $v \sin i$ distributions obtained for the studied sample and in Sect. \ref{conclusions} we present the conclusions.

\section{Observations and the Sample}
\label{observation}

Based on the spectral type as the sole criterion, we selected 379 O9 to B4 main sequence stars from the HIPPARCOS catalogue \citep{Perryman07}. 
High-resolution spectra were then obtained for these stars on January 8, 9
and April 8, 2007 with the MIKE spectrograph at the Magellan Clay 6.5 m telescope on  Las Campanas observatory in Chile. MIKE \citep{Bernstein03} is a double \'echelle spectrograph that registers the whole spectrum on two CCDs (red side $\lambda 4900 - 9500$ \AA, and blue side $\lambda 3350-5000$ \AA) in a single exposure. Here, the blue spectra are analyzed as these contain most of the diagnostic spectral lines needed for estimating $v\sin i$, spectral type, and the effective temperature ($T_{eff}$) of the star. The spectral resolution of the observed spectra is $R\sim55,000$, and were obtained using a slit width of 0.7 arcsec.

In order to minimize possible evolutionary effects  on the $v\sin i$ and given that the He {\sc i} line width calibration adopted in this study (\citealt{daflon07}, Section 4) is valid for main sequence stars, we screened the observed spectra in order to exclude all evolved stars from the sample. 
The Balmer lines and other spectral features which are sensitive to surface gravity such as,
the line ratios $\lambda 4686$ He {\sc ii}/$\lambda 4713$ He {\sc ii}  (stars with spectral types O9--B0),  
and $\lambda 4552$ Si {\sc iii}/$\lambda4387$ He {\sc i}
(stars classified as B1 or later), were used as the primary luminosity criteria.
Our final sample consists of 350 stars and is expected to contain only main sequence stars and not giants or supergiants.

\begin{figure}
\plotone{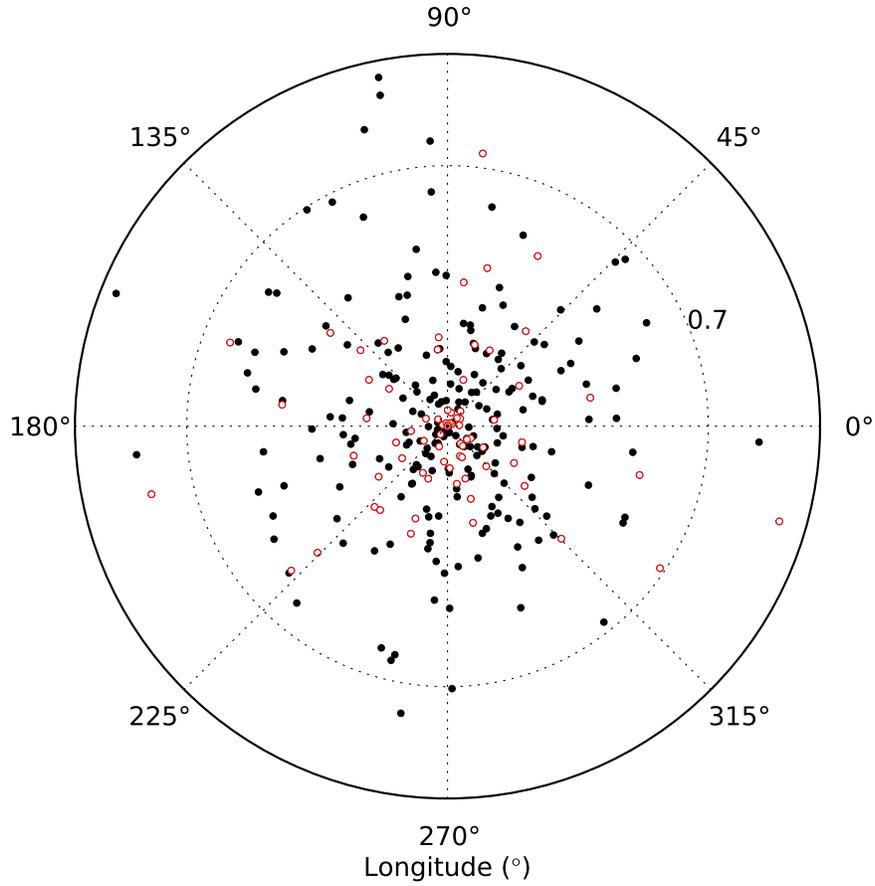}
\caption{Polar plot showing the positions of the sample stars projected onto  the Galactic plane.
The radius is limited to 1 kpc and the concentric dotted circle represents the distance of 0.7 Kpc, 
within which  $\sim$80\% of the stars in our sample are located.
The open red circles are spectroscopic binaries/multiple systems identified in our sample.
 Values for distance of the stars are more uncertain beyond 0.5 kpc of the Sun.}
\label{fig:coords}
\end{figure}

\begin{figure}
\plotone{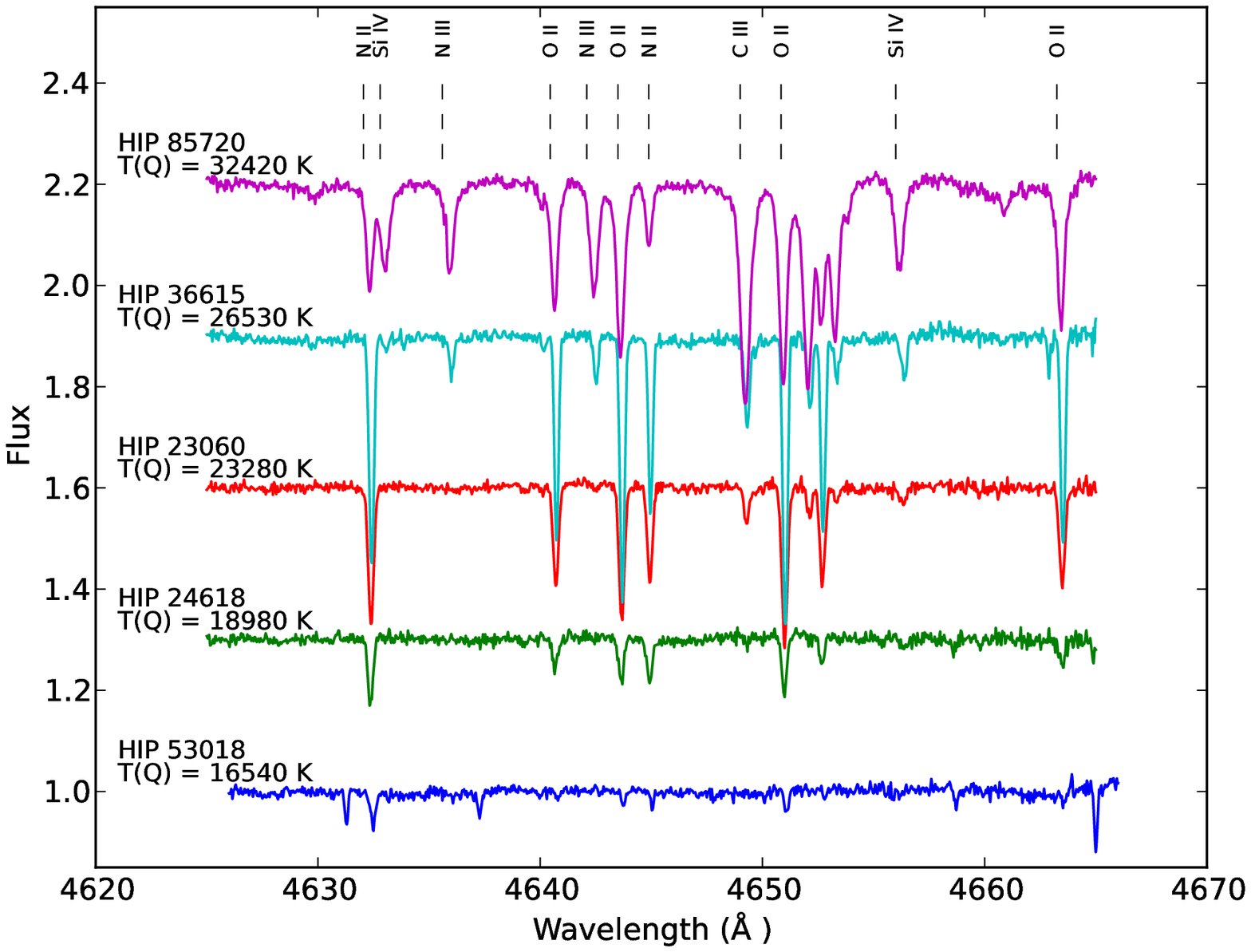}
\caption{Example spectra of five sample stars in the region $\lambda\lambda 4625-4665$ \AA. Some spectral lines are identified.
The spectra were arbitrarily displaced in intensity for better viewing. 
\label{fig:spec}}
\end{figure}

The observed sample of stars is displayed in Fig. \ref{fig:coords} in terms of their Galactic longitude and heliocentric distance projected onto the Galactic plane. 
The stars in the sample are all nearby ($\sim80\%$ is within 700 pc) and relatively bright ($V\sim5-10$). Spectra with signal-to-noise ratios of the order of 100 were achieved with short exposure times ranging from a few seconds to a few minutes.
The spectra were reduced with the Carnegie Observatories python pipeline\footnote{Available at {\tt http://obs.carnegiescience.edu/Code/mike}} and followed standard data reduction procedures: bias subtraction, division by flat field, wavelength
calibration. In addition, small pieces containing the lines of interest were manually normalized to a unit continuum
using the task \texttt{continuum} in IRAF\footnote{http://iraf.noao.edu/}.
Sample spectra are shown in Fig. \ref{fig:spec} in the spectral region between $\lambda\lambda4625-4665$ \AA, which contains spectral lines of C, N, O and Si. The spectra are shown for 5 target stars and these are displayed in order of increasing temperature.

\section{Stellar Characterization}
\label{char}
\subsection{Identification of Spectroscopic Binaries}
\label{binarity}

It is likely that most massive OB stars form in clusters or associations, with the probability of a star forming with a companion being high. 
The recent study by \cite{oudmaijer10}, for example, found a binary fraction of $\sim30\%$ in their photometric survey of B and Be stars. 
A first objective in this study is to identify those stars, among the 350 stars observed, that show spectral signatures of binary or multiple components.
This was done through a careful visual inspection of their spectra. Single-line spectroscopic binaries are 
not detected here, as the spectra are only from single epoch observations.
Spectroscopic binaries will be discarded from further analysis in this study since the  methodology here is most 
appropriate for spectra showing a single component. 

Some stars in our sample were identified as clearly having double, multiple or asymmetric spectral lines.
In addition, some stars in our sample which were  found to be binary or multiple systems in the large survey of stellar multiplicity within the HIPPARCOS catalogue by \cite{eggleton08} and/or appeared as binaries in the study of OB star variability based on HIPPARCOS photometry by \cite{lefevre09}. 
Table \ref{tab:bin} lists 78 stars culled from the sample as spectroscopic binaries or multiple systems, 
representing  22\% of the stars in our sample. Column 1 has the star identification, column 2 lists the spectral types from SIMBAD\footnote{http://simbad.u-strasbg.fr}. In column 3 stars are classified as `SB', if they were found here to be a spectroscopic binary or multiple system and `asym', if they exhibited asymmetric line profiles; ET08 and Lef09 if they were in  \cite{eggleton08} or \cite{lefevre09}.
The stars in Table 1 will not be analyzed in the remainder of this paper.

\subsection{Spectral Types and Effective Temperatures}
\label{spectral}

The spectral types of the stars were determined based on the classification system
presented in the Atlas of OB stars by \cite{WaFi90}.
Relative intensities of some key absorption line ratios such as:
$\lambda 4471 $ He {\sc i} / $\lambda 4481 $ Mg {\sc ii};  $\lambda 4630 $ N {\sc ii} / $\lambda 4631 $ Si {\sc iv}; 
$\lambda 4641 $ N {\sc iii} / $\lambda 4643 $ N {\sc ii},  and 
$\lambda 4649 $ C {\sc iii} / $\lambda 4650 $ O {\sc i}  were used to assign spectral types.
In order to map the Walborn \& Fitzpatrick spectral types into our sample, 
a small grid of non-LTE synthetic spectra of two spectral regions, $\lambda\lambda$4450 -- 4490 \AA\
and $\lambda\lambda$4630 -- 4700 \AA\, were computed for $T_{eff}$'s between 15,000 -- 33,000K, logarithmic of the surface gravity
$\log g = 4.0$, and solar composition. The theoretical spectra were
calculated with the codes \texttt{TLUSTY} and \texttt{SYNPLOT} (\citealt{hubeny88,HubenyLanz95}).
The  Walborn \& Fitzpatrick standard star spectra 
were then visually matched to their closest synthetic counterpart in the grid;
spectral types assigned as O9, B0, B1, B2, B3, B4 and B5 were found to correspond to model spectra 
with $T_{eff}$'s of 33,000K; 30,000K; 25,000K; 20,000K; 18,000K; 16,000K and 15,000K, respectively.

Synthetic and observed spectra were then compared by visual inspection in order to assign spectral types for the target stars.
The goal was simply to determine an appropriate spectral type to each star, and not
to match in detail the observed and theoretical spectra in a fine analysis.
Since a fraction of the stars in our sample have spectral lines somewhat blended by rotation, 
synthetic spectra were convolved for $v\sin i$ (in steps of $v\sin i = 50$ km s$^{-1}$)
in order to aid in the assignment of spectral types of broad lined stars. 
Spectral types for the target stars are listed in Table \ref{tab:sample} (column 2). 

Effective temperatures for the stars were estimated from a calibration 
of the classical reddening free parameter $Q$ (\citealt{johnson58}; $Q=(U-B)-X\cdot(B-V)$, where $X=E(U-B)/E(B-V)$). 
In order to estimate $T_{eff}$ for the sample stars in this study we will adopt the $T(Q)$ calibration presented in \cite{massey89} and defined below:

\begin{equation}
\log T_{eff} = 3.994-0.267\cdot Q+0.364\cdot Q^2.
\end{equation}

\begin{figure}
\plotone{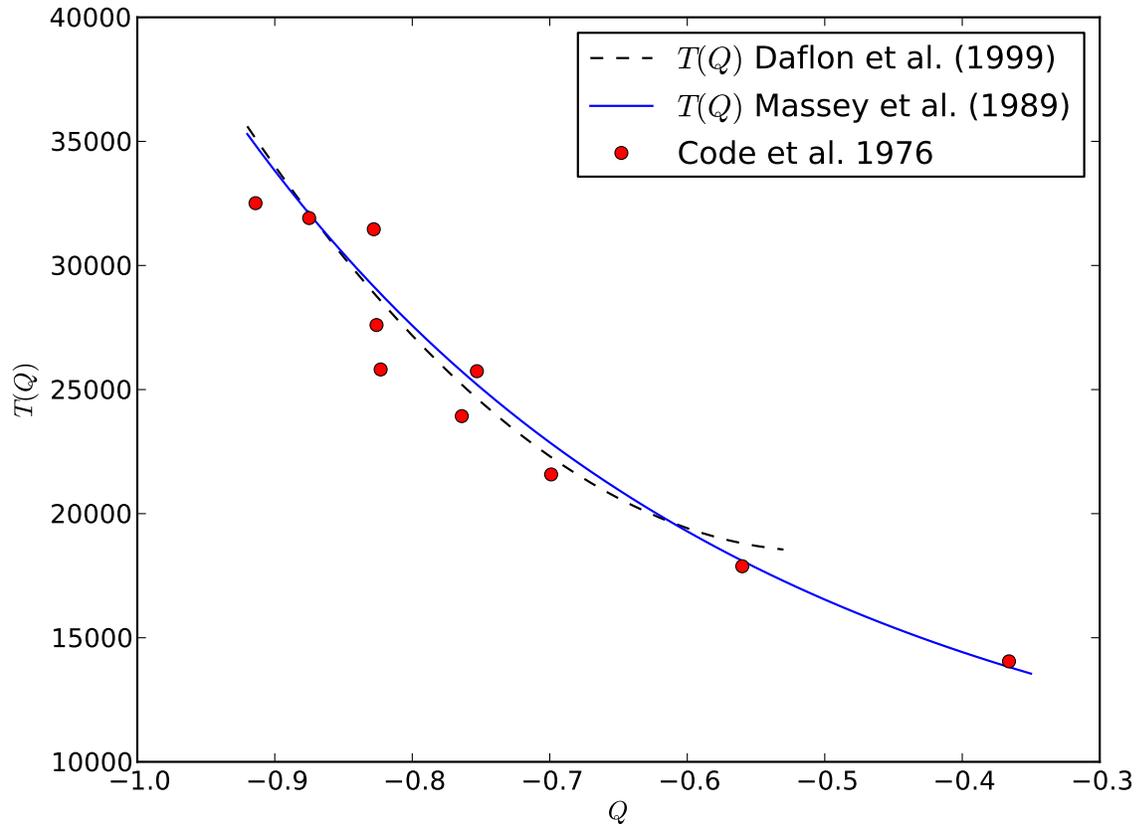}
\caption{The T(Q)  calibration  from \citet[solid blue line]{massey89} which was adopted in this study to estimate effective temperatures for the target stars. The Q-index calibration from \cite{daflon99} is also shown for comparison (black dashed line). 
The red filled circles represent the stars with measured radius and effective temperatures in \cite{code76}. \label{fig:calib}}
\end{figure}

A $T(Q)$ calibration has also been proposed by \cite{daflon99}. However, a large number of stars in the sample studied here are much cooler than the validity range of the Daflon {\it et al.} calibration.
Figure \ref{fig:calib} shows as a solid blue line the calibration by \cite{massey89} for the $Q$-interval of the stars in this study.
The calibration by \cite{daflon99} is also shown in Fig. \ref{fig:calib} as black dashed line, for comparison. 
The average differences between the two calibrations are relatively small:
$\langle\Delta T_{eff}\rangle= -380$ K and $\sigma=  177$ K, for $Q$-values ranging between $-0.62$ and $-0.87$; and $\langle\Delta T_{eff}\rangle= +583$ K and $\sigma = 405$ K for $Q$-values
between $-0.61$ to $-0.53$. 
Effective temperatures for those stars with measured radius from \cite{code76} are shown by red circles in Fig. \ref{fig:calib}.  The overall agreement of the \cite{code76} results with the calibrations is generally good but with significant  scatter, which is indicative  of the uncertainties when using the $Q$-index as a temperature indicator. 
More recently, \cite{paunzen05} also presented a calibration for the $Q$-index with the effective temperature
and the $T\times Q$ relation in that study is quite similar to the one derived in \cite{massey89}. 

The Johnson color indices $(U-B)$ and $(B-V)$ for the studied stars were obtained from \cite{mermilliod87}. For those 57 stars in the sample without published Johnson photometry, UBV colors were computed from Str\"omgren photometry from \cite{HM80, HM98}, using the  transformation in \cite{HB01}.  In addition, there were 41 remaining stars in our sample for which there was no available photometry in the literature, and in those cases we relied on spectral types in order to obtain the intrinsic colors from the tables in \cite{fitzgerald70} and then estimate $Q$.
In columns 3, 4, and 5 of Table 2 we list the $V$ magnitudes, the $Q$ parameters, and the derived $T_{eff}$'s for 272 stars of the observed sample. 
The estimated $T_{eff}$'s here are good for the purpose of a rough stellar characterization of our sample and, in particular, these suffice for a solid derivation of $v\sin i$ values since the grid of synthetic spectra used here (Sect. \ref{velocity}) has been computed for steps of 5,000 K in $T_{eff}$.

\section{Projected Rotational Velocities}
\label{velocity}

\begin{figure}
\plotone{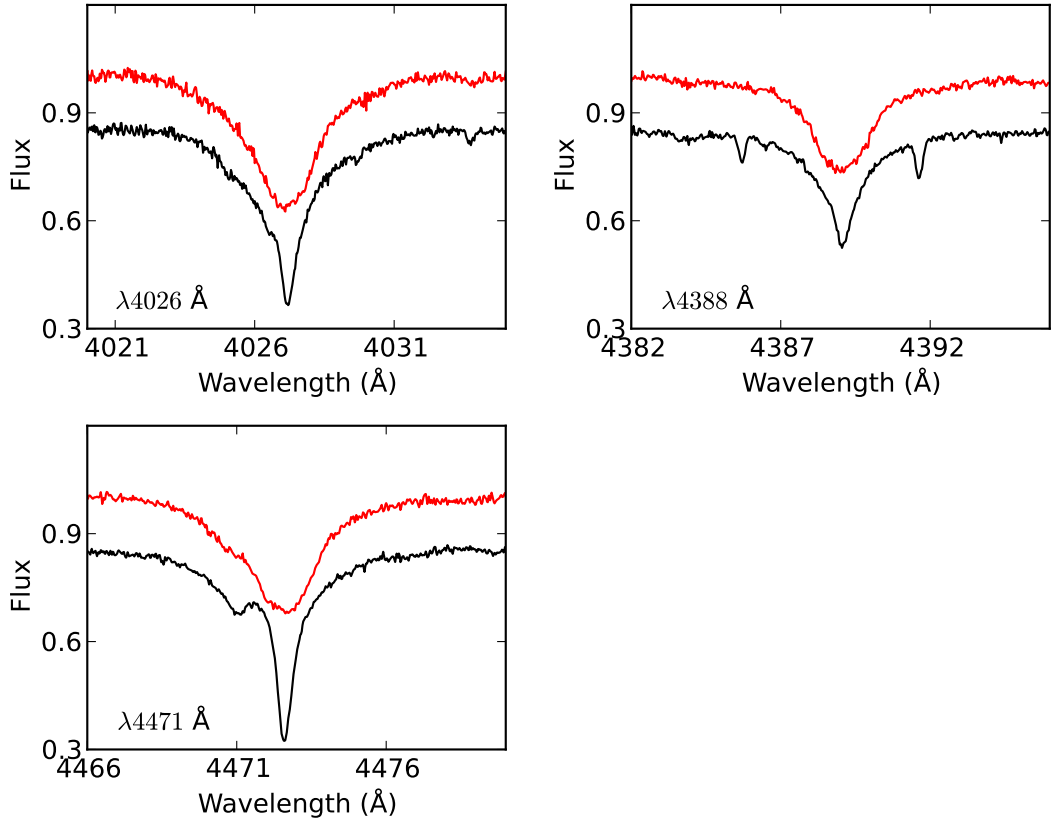}
\caption{Sample spectra showing the 3 He {\sc i} lines that were used to derive the projected rotational velocities for the target stars. The bottom spectra (black) in the three panels are for the star HIP 73624 with $v\sin i = 17 $ km s$^{-1}$ 
and the top spectra (red) are for the star HIP 33492 with $v\sin i = 71$ km s$^{-1}$. 
The spectra were arbitrarily displaced in intensity for better viewing.}
\label{fig:He_lines}
\end{figure}

Projected rotational velocities for the targets were estimated from measurements of the full width at half maximum (FWHM) of 3 He {\sc i} lines at $\lambda4026$ \AA, $\lambda4388$ \AA\ and $\lambda4471$ \AA. The FWHMs of the  He {\sc i} line profiles were measured using the IRAF package \texttt{splot}, using a procedure consistent with that adopted
in \cite{daflon07}: the continuum level was marked at the line center, and the half-width of the red wing was measured at the half-maximum and then doubled in order to derive the FWHM. 
Figure \ref{fig:He_lines} shows examples of the sample He {\sc i} lines for the observed stars HIP 73624 (black continuous line) and HIP 33492 (red dashed line).

The measured FWHM were converted to $v\sin i$'s via interpolating in the grid of synthetic FWHM of He {\sc i} lines presented  in Table 2 of  \cite{daflon07} for the adopted effective temperature of each star. The synthetic He {\sc i} profiles in that study were computed in non-LTE using the codes \texttt{DETAIL} \citep{giddings81}   and \texttt{SURFACE}  \citep{butler85} and were  based on the helium model atom described in \cite{przybilla05}.  We note that the macroturbulent velocity was kept as zero in the calculation of the synthetic profiles by \cite{daflon07} but it is likely to result in additional broadening of the line profiles.
\cite{simon-diaz10} did a careful analysis and disentangled the effects of macroturbulence and rotation in line profiles by using Fourier Transform method and obtained macroturbulent velocities for early B-type dwarfs
that are generally lower than 20 km s$^{-1}$, with a clear trend of
decreasing for late B-types. In order to test the importance of neglecting macroturbulence in the synthetic FWHM of the He lines  we did a test calculation including a gaussian
macroturbulent velocity of 20 km s$^{-1}$. The results indicate that considering the  uncertainties of the method adopted here, including macroturbulence at this level has negligible effect in the measured FWHM of the synthetic spectra of sample He {\sc i} lines.

The measured values of FHWM for the 3 He {\sc i} lines used in the $v\sin i$ determinations are found in Table \ref{tab:sample} (columns 6, 7, and 8); columns 9, 10 and 11 list the $v\sin i$'s for each He line; columns 12 and 13, the final $v\sin i$ values for the studied stars: these represent the average values and the standard deviations in each case.
We note that $v \sin i$'s were not derived for 6 stars with $T_{eff}$'s higher than 33,700 K, as they fell out of the validity of 
the $v \sin i$ calibration from \cite{daflon07}.
 
\begin{figure}
\plotone{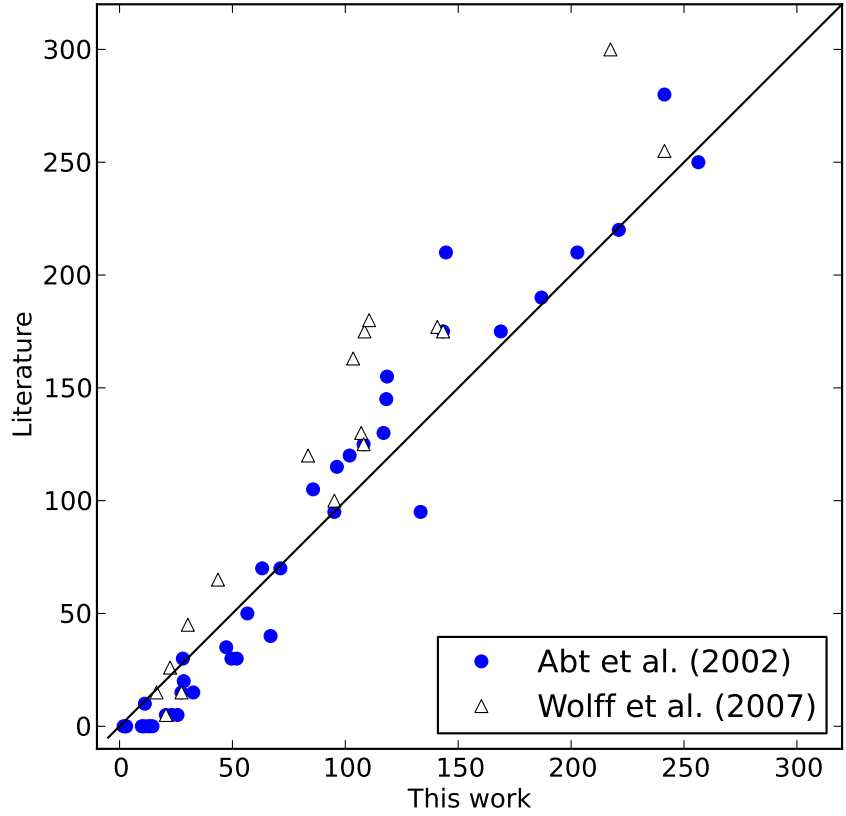}
\caption{A comparison between the $v\sin i$'s derived in this study for stars in common with other two studies in the literature: \cite{ALG02} (blue circles) and   \cite{wolff07} (white triangles). The solid line represents the locus of equal values. 
\label{fig:comparison}}
\end{figure}

Figure \ref{fig:comparison} shows a comparison of the $v\sin i$ results in this study with those from other determinations in the literature: results from \cite{ALG02} are represented as filled blue circles while results from the \cite{wolff07} study are represented as red filled triangles. 
\cite{ALG02} derived $v\sin i$ for a sample of B stars of The Bright Star Catalogue 
with luminosity classes between I and V, using  a calibration for FWHM of He {\sc i} and 
Mg {\sc ii} lines anchored  on standard stars of \cite{Slettebak75}. 
\cite{wolff07} obtained a relationship between FWHM and $v\sin i$ based on  
results from He {\sc i} lines of  \cite{hg06a}.
We note that the $v\sin i$'s for the stars in common with \cite{ALG02} are systematically lower than the ones here in the range between $\sim$ 0 -- 90 km s$^{-1}$ ($\langle\Delta v\sin i \rangle$(This Study -- Abt et al)$ = 9$ km s$^{-1}$ for 24 stars in common); higher than ours in the range between 90 -- 150 km s$^{-1}$ ($\langle\Delta v\sin i\rangle$(This Study -- Abt {\it et al.})$\geq -15$); and in rough agreement for the largest $v \sin i$ (except for one star). The $v\sin i$'s in \cite{wolff07} are mostly higher than ours, except for stars with the lowest $v \sin i$'s. The average $v \sin i$ difference (This study -- Wolff {\it et al.}) is $-26$ km s$^{-1}$ for 17 stars in common. Given the uncertainties in the determinations and the methods adopted, there is reasonable agreement between the three different studies.

\section{Discussion}
\label{discussion}

\subsection {The Entire Sample}

\begin{figure}
\plotone{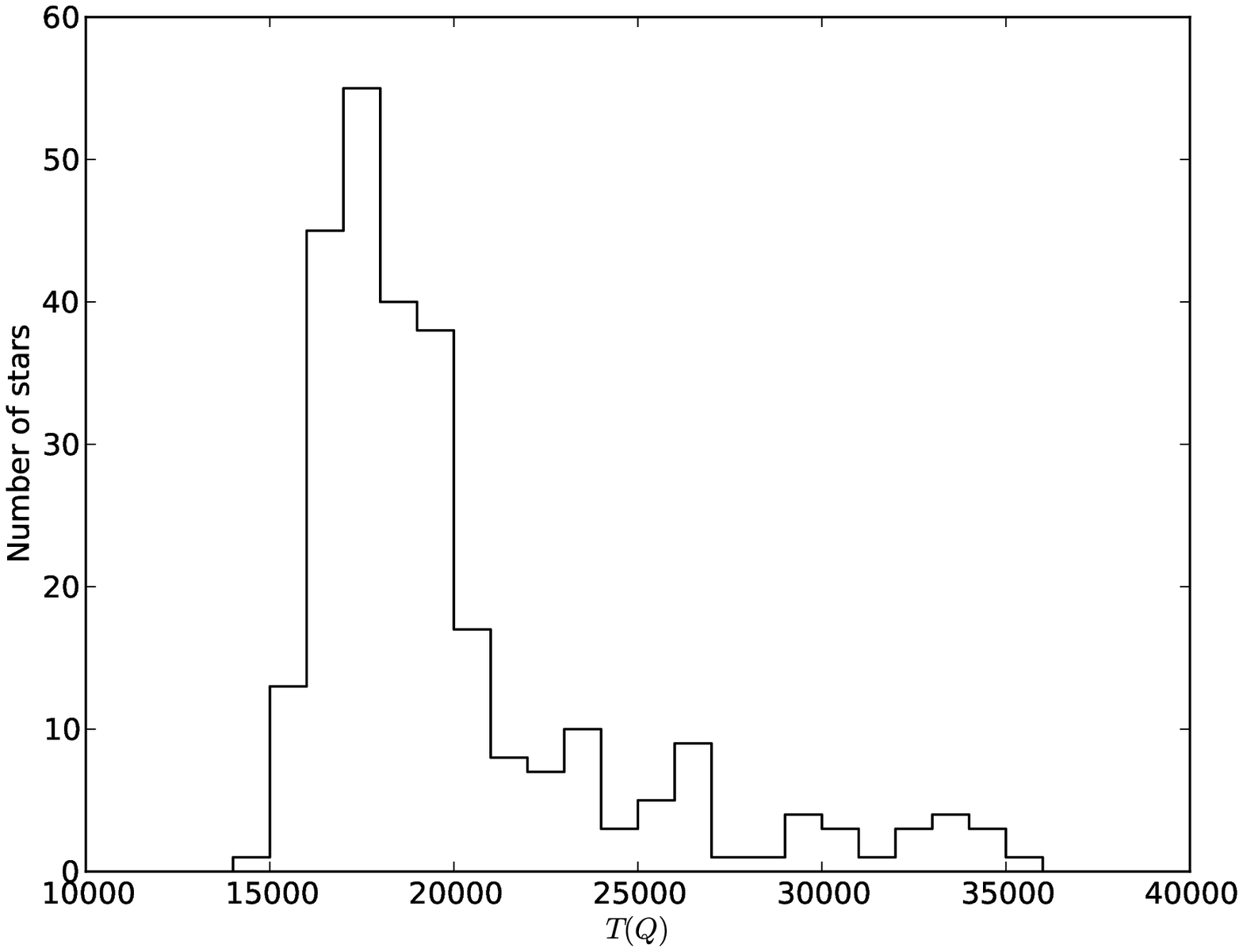}
\caption{Histogram showing the distribution of effective temperatures for the studied sample.}
\label{fig:teff}
\end{figure}

We start our discussion by showing results for the derived effective temperatures for the stars.
A histogram showing the distribution of effective temperatures for 272 OB stars is shown in Fig. \ref{fig:teff}. 
The effective temperatures of the targets sample peak around  17,000K, with most stars being cooler than 28,000K. 

\begin{figure}
\plotone{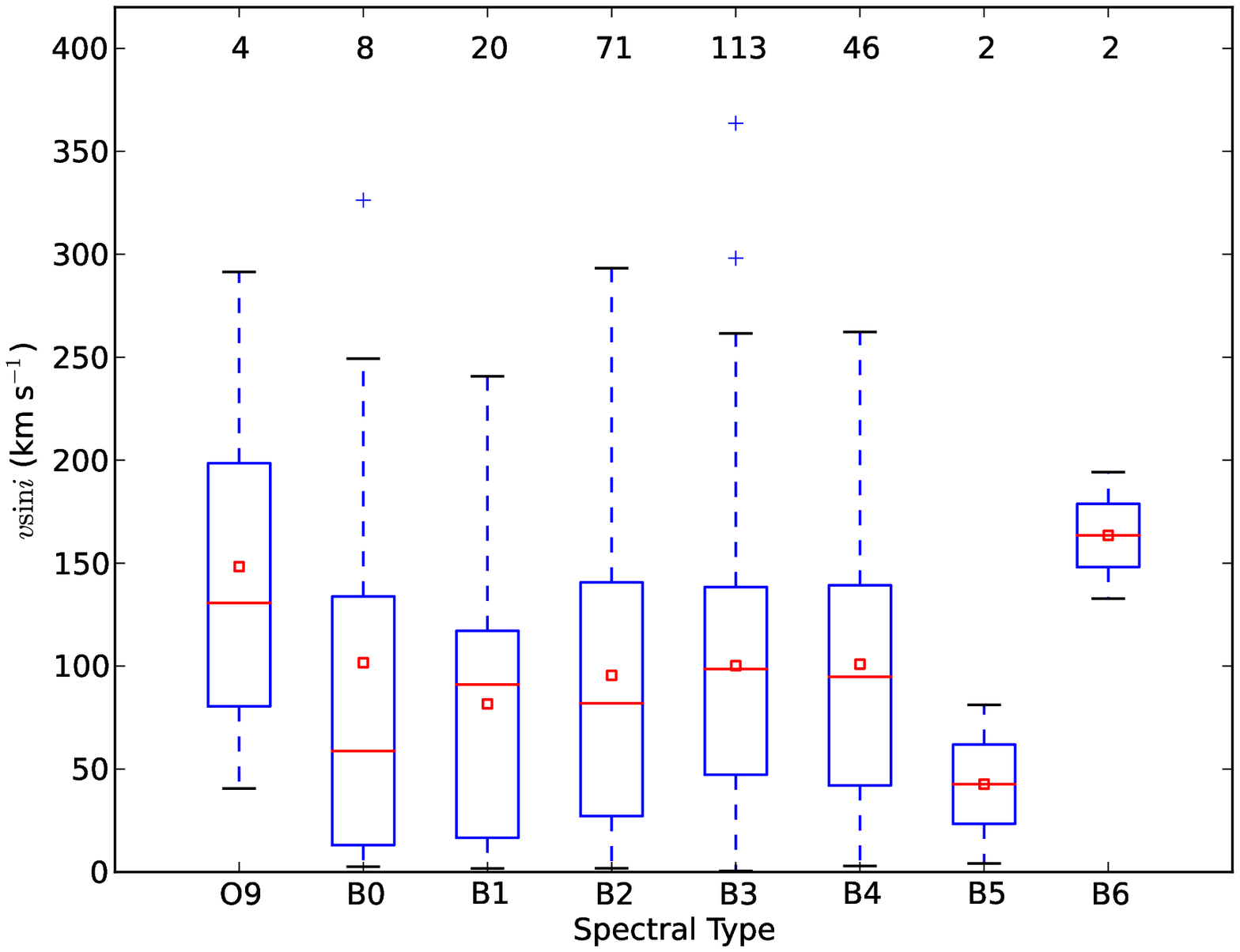}
\caption{Box plot for the studied stars in terms of the spectral type. 
The average $v\sin i$ for  the stars in each spectral type bin is roughly constant, even considering  the least populated bins.
\label{fig:SP}}
\end{figure}

 Figure \ref{fig:SP} illustrates the box plots for the $v\sin i$ values for the  studied stars in each corresponding spectral type.
The box extends from the lower to upper quartile values of the data, with a line at the median and a small box as the mean. The whiskers extend from the box to show the range of the data. The crosses are the outliers.
An inspection of this figure indicates that the mean $v\sin i$ for each spectral type bin is roughly consistent with a constant value across spectral type.
The average $v \sin i$ value computed for the studied sample is  $98$ km s$^{-1}$.
\cite{hg06a} also found a  distribution of mean $v\sin i$ for cluster stars which is basically flat over a similar spectral type  range, although  their study also includes giant stars.
Overall the mean $v\sin i$'s obtained here for spectral types bins B0--B2 and B3--B5 are in rough agreement with the average results for luminosity classes IV and V in \cite{ALG02}.
(see Sect. \ref{velocity} for comparisons the $v\sin i$ for stars in common in the two studies). 

\begin{figure}
\plotone{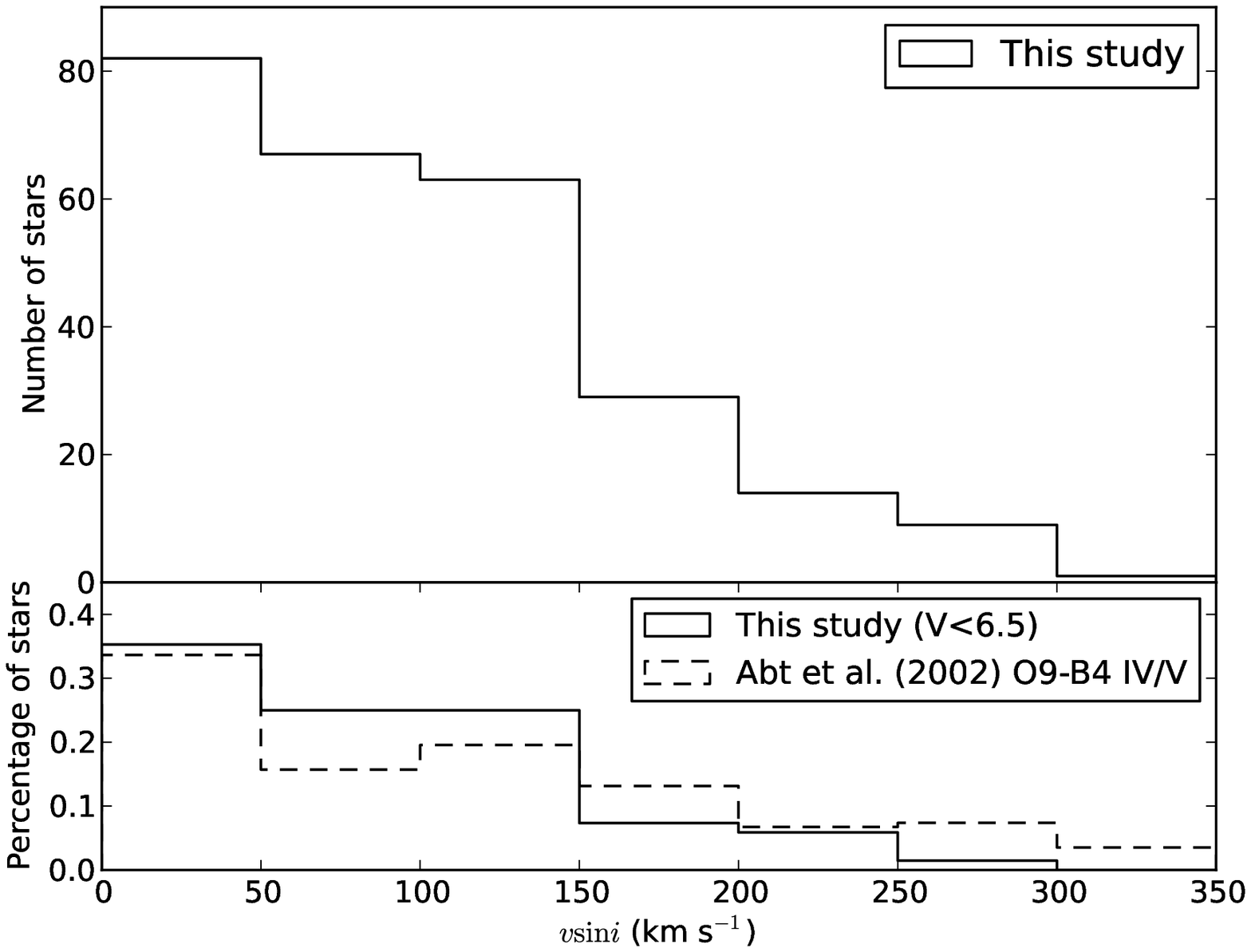}
\caption{Histogram of $v\sin i$ distribution of our sample  on the top panel. The bottom panel compares the normalized distribution of a subsample of  stars in our sample with a magnitude cut in $V=6.5$ and a sample with 312 field stars (spectral types O9--B4IV/V) culled from \cite{ALG02}. 
  \label{fig:Vmean}}
\end{figure}

The $v\sin i$ distribution of the current sample of 266 O and B stars  is shown in  top panel of Fig. \ref{fig:Vmean}.
The distribution has a modest peak at low $v\sin i$'s ($\sim$  0 -- 50 km s$^{-1}$) but it is overall flat (a
broad distribution) for $v\sin i$'s roughly between 0 -- 150 km s$^{-1}$; the number of stars drops for higher values of $v\sin i$. 
As previously mentioned,  the targets in this study were selected considering only their spectral types in the HIPPARCOS catalogue. The sample studied here includes both stars in 
clusters and OB associations, as well as isolated stars that can represent some sort of field population.  

One of the difficulties in making meaningful comparisons between rotational velocity
distributions of stars in clusters versus stars in the `field' is in defining what 
constitutes a `field' star sample. This discussion is, in fact, related to the question 
of whether OB stars can form in isolation and if all OB stars, although isolated, belonged 
in the past to a cluster. The initial idea was that OB stars were only formed in clusters 
and associations but later on were ejected or dispersed into the Galactic field. 
There is growing evidence, however,  that at a least a small fraction of the O 
stars may be born in isolation (or from small molecular clouds). 
For instance, \cite{krumholz09} used a 3-D hydrodynamic simulation to 
show that the formation of isolated massive stars is possible; they successfully form a massive 
binary (having 41.5 M$_\odot$ and 29.2 M$_\odot$) from a 100 M$_\odot$ molecular gas. 
One strong observational evidence that field OB stars may form {\it in situ} was presented by \cite{lamb10}, who found very low mass companions around apparently isolated field OB stars in the SMC. Indeed, \cite{OL11} cite several lines
of empirical evidence to suggest that {\it in situ} field massive stars constitute a significant, and perhaps dominant, component of the  field OB star population.

Although samples of field stars are contaminated at some level with stars that are in the field now but were born in dense environments, 
a comparison of the $v\sin i$'s obtained for the entire sample studied here with other samples taken as representative of the field population is of interest. 
\cite{ALG02} provide the cornerstone work of the distributions 
of projected rotational velocities of the so-called field OB stars. 
The targets in that study were taken from The Bright Star Catalogue and also include stars that are members of clusters and associations.
For the sake of comparison with a field sample that is representative of the spectral types and luminosity classes of most of the studied stars, 
we culled from the \cite{ALG02} sample those stars with spectral types O9 -- B4 and luminosity classes IV and V. The distribution
of $v \sin i$'s for this subsample is shown as the dashed line histogram in  the bottom panel of Fig. \ref{fig:Vmean}. 
 We thus selected those stars of our sample with $V<6.5$ which is the magnitude limit of The Bright Star Catalogue (\citealt{HoffleitJaschek82}) and this subsample is also presented in the the bottom panel of Fig. \ref{fig:Vmean}. 
A Kolmogorov-Smirnov test gives more than  90\% of probability that both distributions are  drawn from the same population.
These results suggest that the $v\sin i$ distribution obtained from \cite{ALG02} for the so-called field population is  similar from the $v\sin i$ distribution of our sample  brighter stars.

\subsection{Stars in OB Associations and Clusters}

The idea that stellar rotation of OB stars in clusters relates to cluster density has been put forward in previous studies in the literature.
In particular, comparisons between the $v \sin i$ distributions of stars from clusters, OB associations, or the field have shown that stellar 
members of dense associations or clusters rotate on average faster than member stars of unbound associations or the field (e.g. \citealt{wolff07,daflon07}).  
Previous studies discussing rotational velocity distributions of stars in clusters include \cite{guthrie82,wolff82,wolff07,hg06a,hg08,hg10}.  In general, all these studies confirm that there seems to be real differences between the $v \sin i$ distributions of cluster members when compared to field; there are fewer slow rotators in the clusters when compared to the field, or, the stars in clusters tend to rotate faster.  \cite{guthrie82}, however, found the presence of a bimodality in his $v\sin i$ distribution: the cluster distribution was double peaked with one at $v\sin i<50$ km s$^{-1}$ and the other at $V\sin i\sim225$ km s$^{-1}$. 

A comparative study of the $v \sin i$'s of all stars in our sample in connection with their birth environments (clusters/associations or field) is of interest but firmly establishing membership is a difficult task as detailed and careful membership determinations are beyond the scope of this paper. 
Instead, in this study, we use literature results in order to select a subsample of stars for which there is secure information on their membership. For OB associations, this is based on the list of probable members from the census of OB associations in the Galactic disk from the HIPPARCOS catalogue by de \cite{dezeeuw99}; and in the study of the stellar content of the Orion association by \cite{brown94}. In addition, we searched the target list in \cite{HM84} and found a few more targets to be association members.  The stars in our sample members in higher density environments or clusters were obtained from cross checking the studied sample with the WEBDA open cluster database (\citealt{MP03}). In addition, we searched the open cluster member list of \cite{robichon99}. The membership information for each star can be found on column 15 of Table \ref{tab:sample}.

\begin{figure}
\plotone{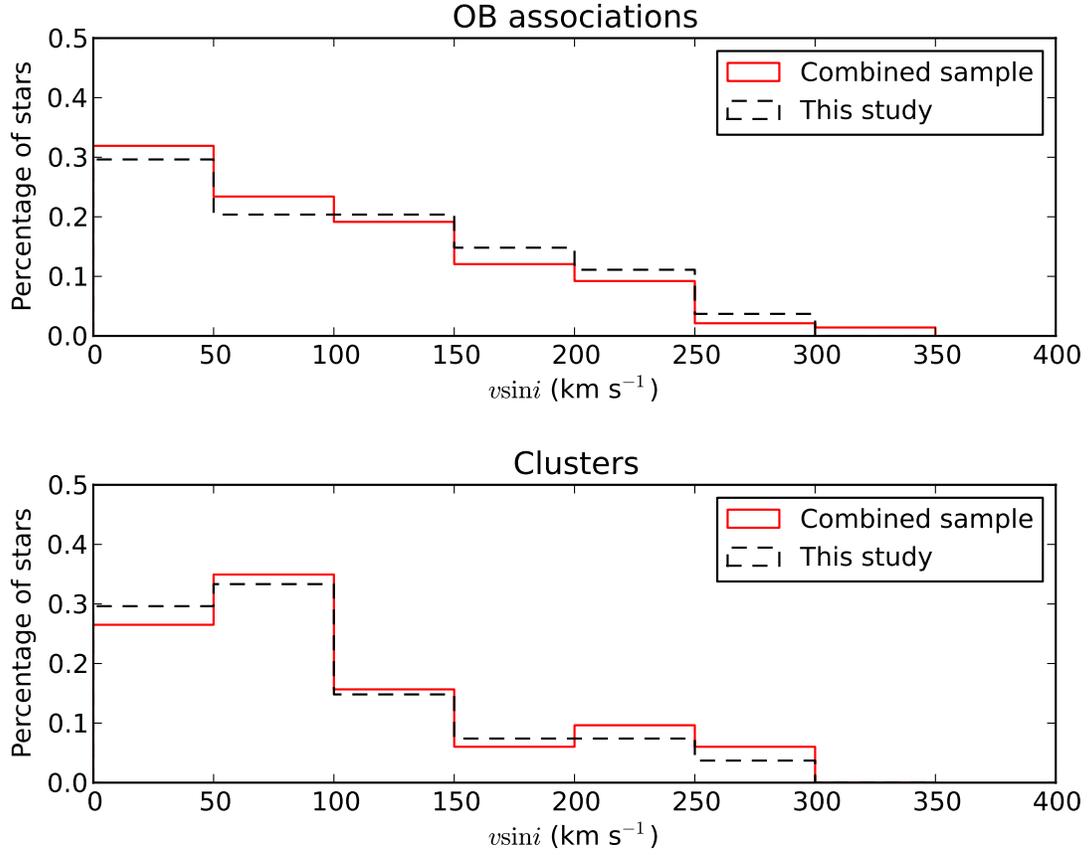}
\caption{Distribution of $v\sin i$'s for the studied samples of OB association (top panel) and cluster members (lower panel) are shown as red dashed lined histograms. The black solid line histograms represent the combined sample: stars in this study plus 
143 star members of clusters and associations from \cite{daflon07}. Both studies use the same methodology to derive $v \sin i$.  \label{fig:A_C}}
\end{figure}

Histograms showing the $v\sin i$ distributions for the culled subsamples of OB association and cluster members are shown in Fig. \ref{fig:A_C} (red dashed lined histograms). The black solid histograms represent a larger sample combining our sample with the sample of O and B stars from \cite{daflon07}. In that paper, 143 OB stars  members of open cluster, OB associations and 23 stars in H {\sc ii} regions have been observed in order to probe the radial metallicity gradient in Galactic disk. Since the $v \sin i$'s in the present study  were derived using the same grid and methodology as in \cite{daflon07}, the discussion beyond this point will be based on the combined sample (black solid histograms) given better statistics.  The distribution of $v\sin i$'s obtained for the stars in OB associations (top panel) has a relatively larger number of objects with $v \sin i$'s between 0 -- 50 km s$^{-1}$ and the number of stars declines smoothly with $v \sin i$. For stars in clusters (bottom panel)  there is a smaller fraction of slow rotating stars and an apparent peak at 50 --100 km s$^{-1}$.  
The smooth distribution of $v \sin i$ values for the association members may result from a nearly single values for equatorial rotational velocity that is viewed at random inclination, while the cluster distribution may be more complex.

\begin{figure}
\plotone{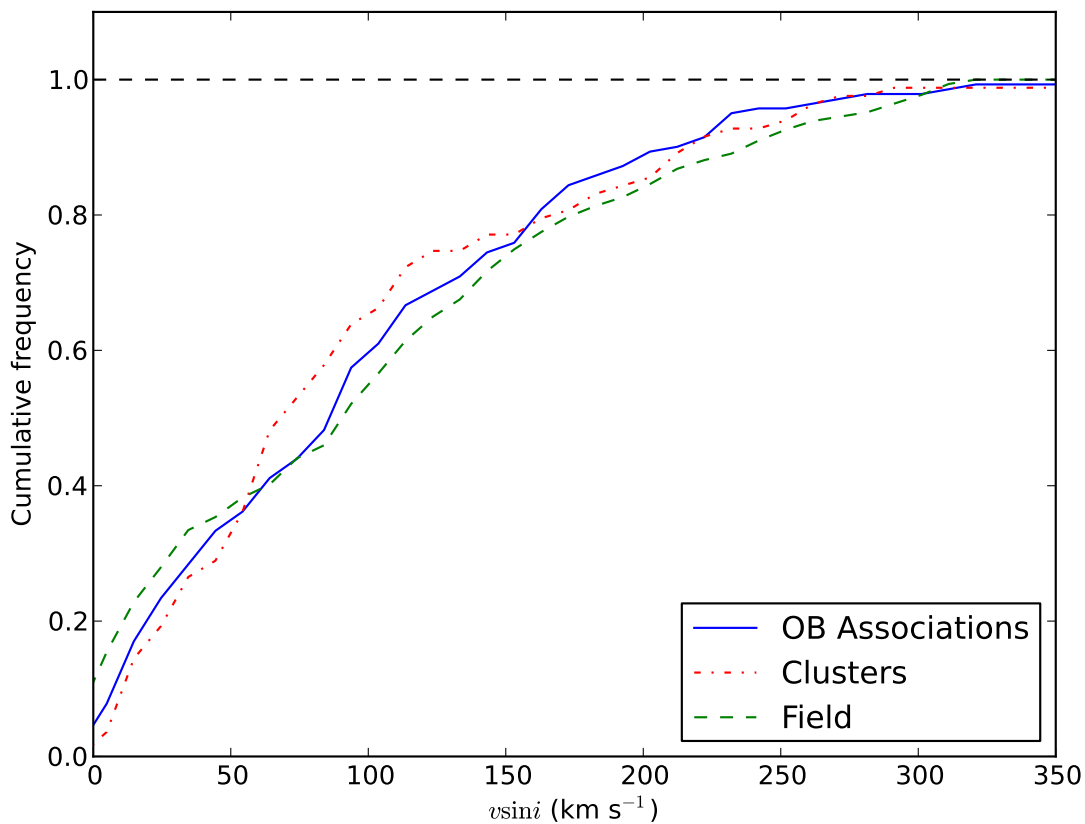}
\caption{Cumulative fractions for the $v \sin i$ distributions for the clusters, OB associations and the field. There seems to be a gradation from cluster to OB association to field. A K-S test between the field star sample and the association sample gives 92 percent probability that both samples are drawn from distinct populations and  88 percent probability that the cluster and the field are drawn from distinct populations. A K-S test between the OB associations and the clusters distributions, however, gives only a 50 percent probability that these are drawn from distinct populations.  \label{fig:cdf}}
\end{figure}

Figure \ref{fig:cdf} shows a comparison of the cumulative fractions for the $v \sin i$ distributions for the clusters and OB associations, as well as the field (from the subsample selected here from \cite{ALG02} as discussed above). The field sample has a higher fraction of slowly rotating stars ($v \sin i$ between 0 and 50 km s$^{-1}$) when compared to the OB associations or clusters. In addition, there is a clear excess of stars with $v \sin i$'s between roughly 70 -- 130 km s$^{-1}$ in the cluster distribution when compared to the OB associations as well as the field. In fact, there seems to be a gradation from cluster to OB association to field confirming the trend found by \cite{wolff07}. A Kolmogorov-Smirnov test between the field star sample and the association sample gives 92 percent probability that both samples are drawn from distinct populations and  88 percent probability that the cluster and the field are drawn from distinct populations. A K-S test between the OB associations and the clusters distributions, however, gives only a 50 percent probability that these are drawn from distinct populations. Thus, any differences between the distributions of clusters and associations in this study are not so clear and may not be statistically significant; larger studies are needed.

\subsection{Runaway Stars}

Few studies in the literature have investigated the distribution of rotational velocities in runaway OB stars.
\cite{martin06} studied the properties of a population of stars far from the Galactic plane and this included a sample of  21 Population I runaway stars. The $v \sin i$ distribution for the runaway stars was found in that study to be broad with no apparent peaks in the range $v\sin i =50$ to 200 km s$^{-1}$ and with a slight decline for values of $v\sin i$ below 50 km s$^{-1}$ (see \citealt{martin06}, Fig. 9b).
The interpretation was that the projected rotational velocity distribution for the runaways was more similar to that of an OB association than to the field; one of the main distinctions when comparing with the field is the absence of a larger number of slow rotators in the distribution of the runaway sample.

Runaway stars can be explained by two scenarios: the binary supernova scenario,  
in which a star is ejected from the binary system when its companion turns into a supernova, and the dynamical ejection scenario,
in which a star is ejected from its parent cluster or association due to dynamical processes.  
These objects are usually identified via one of three methods: spatial velocities, 
tangential velocities, 
or radial velocities.  
\cite{tetzlaff11} combined these three methods to identify runaway stars in the HIPPARCOS catalogue.  
Our study has 34 stars identified as runaways in Tetzlaff {\it et al.}'s catalogue of  runaway candidates. 

\begin{figure}
\plotone{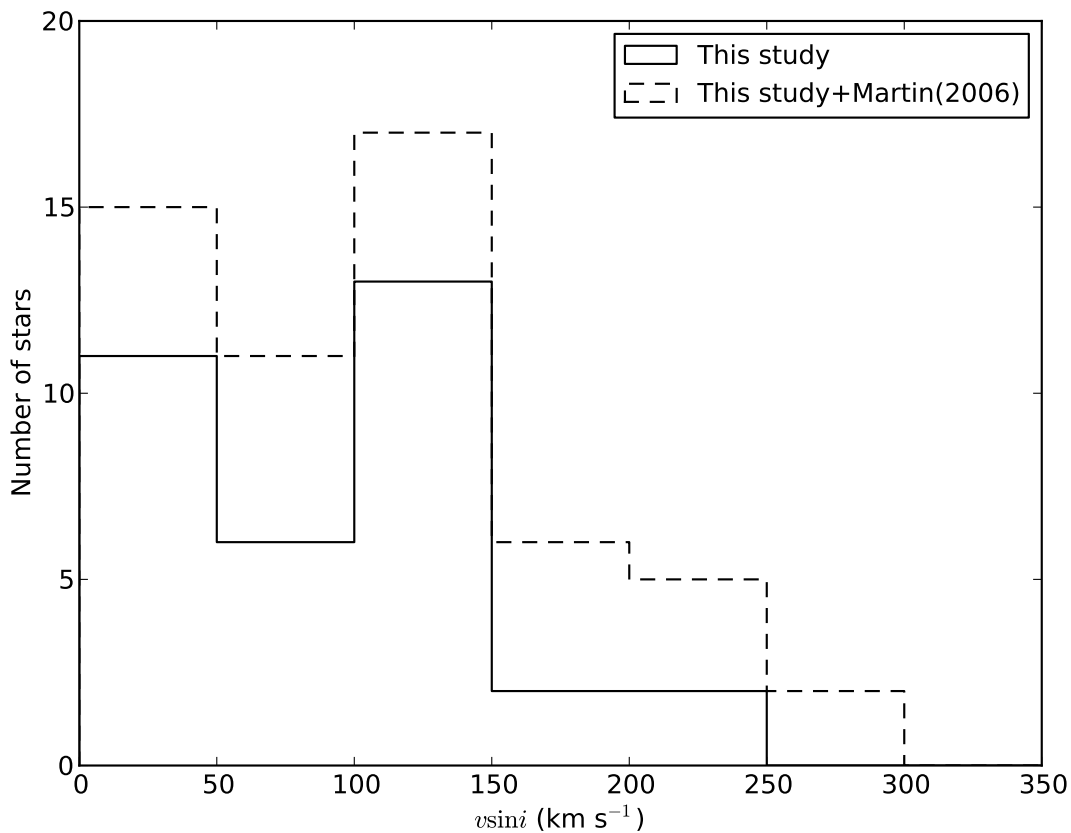}
\caption{$V\sin i$ distribution for the runaway stars in our sample  is shown (solid line histogram). The distribution has two peaks. A KS-test indicates that the runaway $v\sin i$ distribution is more similar to the cluster distribution. This could be an indication that the runaway stars originated from a dynamical ejection scenario. The presence of a second peak at low $v \sin i$  could be related to runaways ejected from OB associations. 
 A histogram representing the combined sample  including the runaway stars studied by \cite{martin06} is also presented for comparison (dashed line histogram).
\label{fig:run}}
\end{figure}

The $v \sin i$ distribution obtained for the runaway stars in our sample  is shown as a solid line histogram  in Fig. \ref{fig:run}. 
Two peaks are evident from a visual inspection of  our distribution: one corresponding to slow rotating stars (or, $v \sin i\sim$  0 -- 50 km s$^{-1}$) and another corresponding to higher projected rotational velocities ($v \sin i$'s between 100 -- 150 km s$^{-1}$). 
 We also show for comparison a histogram representing the combined sample  including the runaway stars studied by \cite{martin06}.
 Given that the distribution of $v\sin i$ in \cite{martin06} runaway sample is generally flat, the two $v\sin i$ peaks observed in the solid line histogram remain in the combined sample.

A K-S test was run on the runaway $v\sin i$ distribution  obtained in this study compared to the other 3 samples discussed previously: the field, the OB association and the cluster subsamples. The probabilities  that both distributions are drawn from the same populations are 18 percent, 40 percent and 71 percent, respectively for the field, association and cluster. This is an indication that the runaway phenomenon maybe more likely associated with the dense cluster environments, as expected from a dynamical ejection scenario. 

However, we note the lack of very massive and dense clusters nearby the Sun, which are the main sources of runaways ejected by means of the dynamical ejection scenario.

As a final note, the presence of a second peak at low $v \sin i$ ($\sim$ 0 --50 km s$^{-1}$) in the runaway distribution in this study,  could be related to runaways originating from OB associations. As discussed previously, stars in associations have tipically lower $v \sin i$ when compared to cluster stars. 

\section{Conclusions}
\label{conclusions}

High resolution spectroscopic observations and a first characterization of a sample of 350 OB stars has been carried out.
Projected rotational velocities were obtained for 266 stars (after rejecting spectroscopic binaries/multiple systems) using measurements of FWHM of He {\sc i} lines and interpolation in a synthetic grid from \cite{daflon07}. The $v\sin i$ distribution obtained for the studied sample has a modest peak at low $v\sin i$'s ($\sim$  0 -- 50 km s$^{-1}$) but it is overall flat for $v\sin i$'s roughly between 0 -- 150 km s$^{-1}$; the number of stars drops for higher values of $v\sin i$.  The $v\sin i$ distribution of our brighter sample stars is similar to the  one obtained from a sample of field stars picked from the work of \cite{ALG02}.

Literature results on membership were used in order to identify subsamples of stars belonging to OB associations or clusters.
We compared these two groups and found that stars members  of OB associations and clusters compose two distinct populations.
The cluster stars tend to have higher $v\sin i$'s when compared to the OB association subsample which could mean that stellar rotation of a population is dictated by the density of the cloud in which it forms.
Also, when the OB association and cluster populations are compared with the field sample, it is found that the latter has a larger fraction of slowest rotators, as previously shown by other works. In fact, there seems to be a gradation from cluster to OB association to field in $v \sin i$ distribution. 

The present sample has 34 stars that were identified as runaway candidates in \cite{tetzlaff11} catalogue.  The $v\sin i$ distribution of the runaways sample 
presents two peaks: one for $v \sin i \sim$ 0 -- 50 km s$^{-1}$ and another for $v \sin i\sim$ 100 -- 150 km s$^{-1}$. The K-S test run with the runaway stars, OB association, cluster and field samples indicate that the runaway $v \sin i$ distribution is more likely to be similar with the distribution of the denser environments,
which could suggest that these stars were ejected through the dynamic ejection mechanism.
Also, there is a possibility that the low $v \sin i$ peak is composed of stars that were ejected from OB associations. 
 
\acknowledgments

We thank the referee for the careful reading and suggestions.
We warmly thank Marcelo Borges, Catherine Garmany, John Glaspey and Joel Lamb for fruitful discussion and comments on the manuscript. G.A.B. thanks the hospitality of University of Michigan on his visit and acknowledges Conselho Nacional de Desenvolvimento Cient\'ifico e Tecnol\'ogico (CNPq-Brazil) and Coordena\c c\~ao de Aperfei\c coamento de Pessoal de N\'ivel Superior (CAPES - Brazil) for his fellowship.  T.B. was funded by grant No. 621-2009-3911 from the Swedish Research Council (VR). M.S.O. and T.B. were supported in part by NSF-AST0448900.  M.S.O. warmly thanks NOAO for the hospitality of a sabbatical visit. 
K.C. acknowledges funding from NSF grant AST-907873. This research has made use of the SIMBAD database, operated at CDS, Strasbourg, France.

{\it Facilities:} \facility{Magellan Observatory}

\clearpage

\input{table1.tex}

\clearpage

\input{table2.tex}

\end{document}

%% file: table1.tex
\begin{deluxetable}{ccl}

\tabletypesize{\scriptsize}
\tablecaption{  \label{tab:bin}}
\tablewidth{0pt}
\tablehead{
\colhead{HIP} & \colhead{Spec. Type} &  \colhead{Bin.} 
}
\startdata
17563	&	 B3V 	&	 Lef09, ET08  	\\
17771	&	  B3V  	&	    ET08  	\\
21575	&	  B3V  	&	  SB, Lef09   	\\
22663	&	  B2/B3V  	&	  Asym., Lef09   	\\
25028	&	  B3V  	&	  SB,  ET08  	\\
25066	&	  B3V  	&	  SB    	\\
25142	&	  B2  	&	  SB,  ET08  	\\
26063	&	  B3V  	&	   Lef09, ET08  	\\
26213	&	  B3V  	&	  SB    	\\
28142	&	  B2V  	&	  SB?, Lef09   	\\
28756	&	  B2.5V  	&	   Lef09, ET08  	\\
29126	&	  B1V  	&	  asym:    	\\
29321	&	  B2V  	&	  SB, Lef09   	\\
31068	&	  B3V  	&	  SB, Lef09   	\\
31190	&	  B2IV  	&	  SB, Lef09, ET08  	\\
31593	&	  B3V  	&	  Asym.?, Lef09   	\\
31959	&	  B3V  	&	  SB    	\\
32810	&	  B6Vnpe  	&	  SB, Lef09, ET08  	\\
33330	&	  B3V  	&	  SB    	\\
33723	&	  O9V  	&	  SB    	\\
33971	&	  B1V  	&	   Lef09, ET08  	\\
34579	&	  B2V  	&	   Lef09, ET08  	\\
34817	&	  B3V  	&	  Asym.,  ET08  	\\
35202	&	  B4V  	&	  SB,  ET08  	\\
35611	&	  B2.5V  	&	  Asym., Lef09, ET08  	\\
35621	&	  B2V  	&	  asym.    	\\
36363	&	  B5Vp  	&	  SB,  ET08  	\\
37450	&	  B5V  	&	  SB,  ET08  	\\
37502	&	  B2V  	&	  SB    	\\
37524	&	  B5V  	&	  SB?    	\\
37668	&	  B4V  	&	  SB    	\\
37784	&	  B2V  	&	  SB    	\\
38020	&	  B1.5IV  	&	  SB,  ET08  	\\
38455	&	  B2.5V  	&	   Lef09, ET08  	\\
39943	&	  B4V  	&	    ET08  	\\
39992	&	  B4V  	&	  SB    	\\
40341	&	  B1.5V  	&	  SB    	\\
40443	&	  B4V  	&	  SB    	\\
41039	&	  B1V  	&	  SB  ET08  	\\
41250	&	  B3V  	&	  SB, Lef09, ET08	\\
41296	&	  B2Ve  	&	    ET08 	\\
41515	&	  B3V  	&	  SB, Lef09, ET08 	\\
41621	&	  B2Vne  	&	  SB  ET08 	\\
43955	&	  B2IV  	&	  asym.   	\\
46296	&	  B2/B3IV  	&	  SB   	\\
47559	&	  B4V  	&	  SB,  ET08 	\\
49695	&	  B4V  	&	  SB   	\\
50780	&	  B3V  	&	  SB, Lef09  	\\
52370	&	  B4V  	&	  SB?, Lef09, ET08 	\\
53089	&	  B3V  	&	  SB   	\\
55350	&	  B4V  	&	  SB   	\\
57669	&	  B3Vne  	&	   Lef09, ET08 	\\
59449	&	  B3V  	&	  Asym.,  ET08 	\\
60905	&	  B3IV/V  	&	  SB   	\\
62322	&	  B2.5V  	&	    ET08 	\\
64719	&	  B4V  	&	  SB   	\\
65271	&	  B3V  	&	    ET08 	\\
67042	&	  B3V  	&	  SB   	\\
69648	&	  B3V  	&	  asym.   	\\
74117	&	  B3V  	&	  SB,  ET08 	\\
74680	&	  B3V  	&	  SB   	\\
77811	&	  B3V  	&	  SB,  ET08 	\\
77939	&	  B2/B3V  	&	    ET08 	\\
78004	&	  B3/B4V  	&	  SB   	\\
78168	&	  B3V  	&	  SB?, Lef09, ET08 	\\
78821	&	  B2V  	&	  SB,  ET08 	\\
79404	&	  B2V  	&	    ET08 	\\
80405	&	  B4V  	&	  SB   	\\
88993	&	  B1V  	&	  SB   	\\
90804	&	  B2V  	&	  SB, Lef09, ET08 	\\
91352	&	  B2V  	&	  SB   	\\
92470	&	  B2IV  	&	  asym.   	\\
92989	&	  B8IVs...  	&	  SB, Lef09, ET08 	\\
93225	&	  B4V  	&	  SB  ET08 	\\
93502	&	  B2V  	&	  SB, Lef09  	\\
93732	&	  B3V  	&	  SB, Lef09  	\\
99457	&	  B1V  	&	  SB, Lef09  	\\
100881	&	  B4V  	&	  SB,  ET08 	\\

\enddata
\tablecomments{The columns are: (1) The HIPPARCOS identification, (2) the spectral type, (3) Binarity/multiplicity:
 We classified the star as ``asym." when it has an asymetric line profile or ``SB" when it is a spectroscopic binary.  Some stars are listed as binaries in \cite{lefevre09} (Lef09) and/or
 \cite{eggleton08} (ET08).}
\tablecomments{Table \ref{tab:bin} is published in its entirety in the 
electronic edition of the {\it Astrophysical Journal}. }
\end{deluxetable}

%% file: table2.tex
\begin{deluxetable}{ccccccccrrrrrcc}
\rotate
\tablecolumns{15}
\setlength{\tabcolsep}{0.03in}
\tabletypesize{\scriptsize}
\tablecaption{\label{tab:sample}}
\tablehead{
\colhead{HIP} & \colhead{Spec. Type} & \colhead{$V$} & \colhead{$Q$} & \colhead{$T_{eff}$} & \multicolumn{3}{c}{FWHM} & \multicolumn{3}{c}{$v\sin i$ } & \colhead{$\langle V\sin i\rangle$} & \colhead{$\sigma(V\sin i)$} & \colhead{N} & \colhead{Memb.}\\

\colhead{ } & \colhead{ } & \colhead{ } & \colhead{ } & \colhead{(K)} & \colhead{ } & 
\colhead{(\AA)} & \colhead{ } & \colhead{ } & \colhead{(km s$^{-1}$)} & \colhead{ } & \colhead{(km s$^{-1}$)} & \colhead{(km s$^{-1}$)} & \colhead{ } & \colhead{ }\\

\cline{6-8} \cline{9-11} \\

\colhead{ } & \colhead{ }  & \colhead{ } & \colhead{ } & \colhead{ } & 
\colhead{$\lambda4026$\AA} & \colhead{$\lambda4388$\AA} & \colhead{$\lambda4471$\AA} & \colhead{$\lambda4026$\AA} & \colhead{$\lambda4388$\AA} & \colhead{$\lambda4471$\AA} &\colhead{ } & \colhead{ } & \colhead{ } \\
 
\colhead{(1)} & \colhead{(2)} & \colhead{(3)} & \colhead{(4)} & \colhead{(5)} & \colhead{(6)} & 
\colhead{(7)} & \colhead{(8)} & \colhead{(9)} & \colhead{(10)} & \colhead{(11)} & \colhead{(12)} &
\colhead{(13)} & \colhead{(14)} & \colhead{(15)}
}

\startdata
14898	&	B5V	&	7.03	&	-0.41	&	14610	&	\nodata	&	1.0	&	0.7	&	\nodata	&	5	&	4	&	4	&	1	&	2	&	R	\\
15188	&	B3Ve	&	7.96	&	-0.53	&	17290	&	4.2	&	3.9	&	3.7	&	153	&	151	&	118	&	141	&	19	&	3	&	R	\\
16466	&	B4V	&	9.32	&	-0.51	&	16780	&	1.7	&	1.5	&	1.0	&	22	&	30	&	\nodata	&	26	&	5	&	2	&	R	\\
18926	&	B3V	&	6.45	&	-0.62	&	19930	&	5.7	&	5.7	&	5.4	&	234	&	235	&	194	&	221	&	23	&	3	&	\nodata	\\
18957	&	B3V	&	5.31	&	-0.5	&	16540	&	1.8	&	1.6	&	1.2	&	28	&	33	&	23	&	28	&	5	&	3	&	\nodata	\\
20884	&	B3V	&	5.53	&	-0.48	&	16070	&	1.4	&	1.4	&	0.9	&	15	&	\nodata	&	11	&	13	&	3	&	2	&	A	\\
23060	&	B2V	&	6.13	&	-0.71	&	23280	&	2.1	&	1.3	&	1.1	&	31	&	20	&	\nodata	&	26	&	8	&	2	&	R	\\
23364	&	B3V	&	4.81	&	-0.61	&	19600	&	2.5	&	2.1	&	1.8	&	52	&	60	&	44	&	52	&	8	&	3	&	\nodata	\\
24618	&	B2V	&	6.53	&	-0.59	&	18980	&	1.6	&	1.3	&	1.0	&	6	&	16	&	\nodata	&	11	&	6	&	3	&	\nodata	\\
25368	&	B3V	&	6.42	&	-0.59	&	18980	&	3.0	&	2.7	&	2.6	&	82	&	91	&	77	&	84	&	7	&	3	&	A	\\
25480	&	B4V	&	6.62	&	-0.6	&	19290	&	5.6	&	5.0	&	5.9	&	230	&	203	&	219	&	217	&	14	&	3	&	A	\\
25493	&	B2V	&	6.73	&	-0.6	&	19290	&	3.4	&	3.0	&	3.5	&	100	&	106	&	104	&	103	&	3	&	3	&	A	\\
25496	&	B3V	&	7.19	&	-0.52	&	17030	&	2.2	&	1.9	&	1.5	&	45	&	48	&	38	&	44	&	5	&	3	&	A.R	\\
25582	&	B2V	&	6.38	&	-0.63	&	20260	&	1.4	&	1.4	&	1.1	&	\nodata	&	21	&	\nodata	&	21	&	\nodata	&	1	&	A.R	\\
25751	&	B2V	&	5.74	&	-0.7	&	22870	&	3.2	&	3.0	&	3.6	&	98	&	114	&	112	&	108	&	9	&	3	&	A.R	\\
25844	&	B3V	&	7.63	&	-0.57	&	18380	&	1.9	&	1.8	&	1.3	&	22	&	\nodata	&	22	&	22	&	1	&	2	&	\nodata	\\
25850	&	B2V	&	7.95	&	-0.68	&	22070	&	3.4	&	3.2	&	3.8	&	104	&	123	&	120	&	116	&	10	&	3	&	A	\\
25869	&	B2V	&	6.20	&	-0.61	&	19600	&	2.1	&	1.6	&	1.2	&	23	&	32	&	\nodata	&	27	&	7	&	2	&	A	\\
25881	&	B3V	&	7.56	&	-0.52	&	17030	&	1.8	&	1.8	&	1.4	&	27	&	\nodata	&	33	&	30	&	4	&	2	&	A	\\
25923	&	O9V	&	4.62	&	-0.92	&	35300	&	\nodata	&	\nodata	&	\nodata	&	\nodata	&	\nodata	&	\nodata	&	\nodata	&	\nodata	&	0	&	\nodata	\\
26093	&	B3V	&	5.59	&	-0.53	&	17290	&	3.6	&	3.4	&	3.5	&	118	&	129	&	109	&	118	&	10	&	3	&	C	\\
26098	&	B2V	&	6.57	&	-0.68	&	22070	&	3.9	&	3.8	&	4.3	&	135	&	151	&	145	&	143	&	8	&	3	&	A.R	\\
26166	&	B3V	&	6.71	&	-0.59	&	18980	&	3.4	&	3.8	&	3.7	&	102	&	\nodata	&	115	&	108	&	9	&	2	&	A	\\
26258	&	O9V	&	6.87	&	-0.88	&	32420	&	4.0	&	4.1	&	4.5	&	161	&	173	&	169	&	168	&	6	&	3	&	C	\\
26439	&	B4V	&	7.09	&	-0.49	&	16300	&	3.0	&	2.6	&	2.9	&	93	&	88	&	94	&	92	&	3	&	3	&	A	\\
26508	&	B2V	&	6.81	&	-0.64	&	20600	&	5.7	&	5.5	&	6.5	&	236	&	224	&	248	&	236	&	12	&	3	&	A	\\
26581	&	B3V	&	7.55	&	-0.45	&	15410	&	3.3	&	3.0	&	3.1	&	113	&	108	&	100	&	107	&	7	&	3	&	A.R	\\
26785	&	B2V	&	7.63	&	-0.61	&	19600	&	2.5	&	1.8	&	1.6	&	48	&	47	&	34	&	43	&	8	&	3	&	A	\\
26876	&	B4V	&	8.90	&	-0.54	&	17550	&	5.2	&	5.3	&	5.8	&	203	&	217	&	214	&	211	&	7	&	3	&	A	\\
27103	&	B3V	&	6.76	&	-0.6	&	19290	&	3.0	&	2.8	&	3.4	&	80	&	99	&	100	&	93	&	11	&	3	&	\nodata	\\
27937	&	B3V	&	6.09	&	-0.51	&	16780	&	1.9	&	1.5	&	1.3	&	30	&	28	&	28	&	29	&	1	&	3	&	\nodata	\\
28973	&	B3V	&	6.64	&	-0.56	&	18100	&	2.2	&	1.9	&	1.8	&	41	&	51	&	49	&	47	&	6	&	3	&	A	\\
29121	&	B3Vn	&	8.79	&	-0.64	&	20600	&	8.0	&	8.9	&	8.8	&	355	&	375	&	361	&	364	&	10	&	3	&	C	\\
29127	&	B1V	&	8.33	&	-0.7	&	22870	&	1.9	&	1.4	&	1.3	&	18	&	\nodata	&	17	&	17	&	1	&	2	&	C	\\
29201	&	B0Ve	&	8.58	&	-0.85	&	30480	&	5.3	&	6.0	&	6.4	&	233	&	255	&	261	&	249	&	15	&	3	&	R	\\
29213	&	B4V	&	7.96	&	-0.54	&	17550	&	4.5	&	4.4	&	4.9	&	165	&	174	&	171	&	170	&	5	&	3	&	R	\\
29387	&	B3V	&	8.61	&	-0.6	&	19290	&	4.2	&	4.2	&	4.6	&	150	&	166	&	154	&	157	&	8	&	3	&	\nodata	\\
29417	&	B2V	&	5.05	&	-0.64	&	20600	&	1.6	&	1.0	&	1.0	&	0	&	3	&	\nodata	&	2	&	2	&	2	&	A.R	\\
29429	&	B1V	&	7.05	&	-0.67	&	21690	&	3.1	&	2.5	&	3.0	&	85	&	87	&	88	&	87	&	1	&	3	&	\nodata	\\
29446	&	B2V	&	7.49	&	-0.54	&	17550	&	2.8	&	2.6	&	3.1	&	78	&	87	&	94	&	86	&	8	&	3	&	\nodata	\\
29678	&	B1V	&	5.89	&	-0.79	&	27050	&	1.7	&	1.1	&	1.0	&	11	&	16	&	\nodata	&	14	&	3	&	2	&	R	\\
29771	&	B2/B3Ve	&	5.96	&	-0.63	&	20260	&	6.1	&	6.2	&	6.6	&	257	&	256	&	256	&	256	&	1	&	3	&	\nodata	\\
29941	&	B2/B3V	&	5.49	&	-0.58	&	18680	&	2.6	&	2.2	&	2.6	&	58	&	65	&	78	&	67	&	11	&	3	&	\nodata	\\
30143	&	B1.5Ve	&	5.53	&	-0.76	&	25540	&	5.4	&	5.8	&	6.3	&	228	&	246	&	249	&	241	&	11	&	3	&	R	\\
30382	&	B3V	&	6.38	&	-0.52	&	17030	&	1.4	&	1.2	&	0.8	&	11	&	9	&	\nodata	&	10	&	2	&	2	&	\nodata	\\
30468	&	B4Ve	&	6.12	&	-0.57	&	18380	&	1.9	&	1.6	&	0.8	&	21	&	35	&	\nodata	&	28	&	10	&	2	&	\nodata	\\
30700	&	B4V	&	6.49	&	-0.44	&	15200	&	5.5	&	5.9	&	6.0	&	220	&	246	&	222	&	229	&	15	&	3	&	C	\\
30739	&	B3V	&	8.80	&	-0.6	&	19290	&	3.2	&	3.0	&	3.4	&	90	&	107	&	102	&	99	&	9	&	3	&	\nodata	\\
30743	&	B4V	&	7.49	&	-0.54	&	17550	&	3.1	&	3.0	&	3.0	&	90	&	106	&	91	&	96	&	9	&	3	&	\nodata	\\
30772	&	B2V	&	5.04	&	-0.63	&	20260	&	2.6	&	2.2	&	2.4	&	52	&	69	&	69	&	63	&	10	&	3	&	C	\\
30788	&	B4V	&	4.46	&	-0.49	&	16300	&	3.2	&	2.7	&	3.4	&	102	&	93	&	111	&	102	&	9	&	3	&	\nodata	\\
31028	&	B4Ve	&	7.20	&	-0.54	&	17550	&	3.4	&	3.1	&	2.9	&	110	&	111	&	91	&	104	&	12	&	3	&	\nodata	\\
31106	&	B0.2V	&	7.95	&	-0.85	&	30480	&	2.6	&	2.8	&	3.0	&	80	&	109	&	97	&	95	&	15	&	3	&	C	\\
31305	&	B0V	&	7.65	&	-0.87	&	31750	&	2.3	&	2.5	&	2.9	&	\nodata	&	97	&	94	&	95	&	2	&	3	&	C	\\
31411	&	B1V	&	8.04	&	-0.83	&	29270	&	3.2	&	3.2	&	3.7	&	114	&	126	&	128	&	123	&	8	&	3	&	\nodata	\\
31436	&	B2/B3V	&	7.85	&	-0.66	&	21320	&	6.6	&	6.4	&	6.8	&	284	&	264	&	265	&	271	&	11	&	3	&	A	\\
31824	&	B3V	&	9.69	&	-0.62	&	19930	&	2.1	&	1.6	&	1.4	&	24	&	\nodata	&	18	&	21	&	4	&	2	&	\nodata	\\
31875	&	B3V	&	7.41	&	-0.56	&	18100	&	4.2	&	3.9	&	4.1	&	150	&	151	&	132	&	144	&	11	&	3	&	R	\\
31955	&	B2V	&	7.95	&	-0.74	&	24600	&	2.6	&	2.2	&	2.1	&	66	&	74	&	63	&	67	&	6	&	3	&	C	\\
32007	&	B4V	&	6.12	&	-0.49	&	16300	&	4.3	&	4.4	&	4.9	&	158	&	173	&	171	&	167	&	8	&	3	&	\nodata	\\
32084	&	B3V	&	8.09	&	-0.54	&	17550	&	3.8	&	3.6	&	3.7	&	127	&	135	&	115	&	126	&	10	&	3	&	A	\\
32112	&	B3V	&	6.90	&	-0.46	&	15630	&	2.4	&	1.9	&	1.8	&	67	&	52	&	58	&	59	&	8	&	3	&	\nodata	\\
32454	&	B4V	&	8.24	&	-0.54	&	17550	&	5.2	&	5.9	&	6.1	&	204	&	242	&	228	&	225	&	19	&	3	&	\nodata	\\
33007	&	B4V	&	7.43	&	-0.49	&	16300	&	4.3	&	4.6	&	4.7	&	157	&	182	&	160	&	166	&	14	&	3	&	A	\\
33182	&	B2/B3V	&	8.80	&	-0.51	&	16780	&	1.4	&	1.2	&	0.9	&	11	&	9	&	9	&	10	&	1	&	3	&	\nodata	\\
33208	&	B3V	&	8.84	&	-0.52	&	17030	&	4.8	&	4.1	&	4.8	&	183	&	160	&	165	&	169	&	12	&	3	&	A	\\
33211	&	B3V	&	9.14	&	-0.56	&	18100	&	2.6	&	2.3	&	2.2	&	62	&	74	&	65	&	67	&	6	&	3	&	A	\\
33288	&	B3V	&	8.23	&	-0.57	&	18380	&	3.8	&	3.5	&	3.4	&	125	&	135	&	104	&	121	&	15	&	3	&	\nodata	\\
33457	&	B3V	&	9.58	&	-0.54	&	17550	&	3.7	&	3.5	&	3.8	&	122	&	131	&	123	&	125	&	5	&	3	&	\nodata	\\
33490	&	B3V	&	7.98	&	-0.53	&	17290	&	3.7	&	3.3	&	3.6	&	122	&	121	&	113	&	119	&	5	&	3	&	R	\\
33492	&	B2.5V	&	6.23	&	-0.57	&	18380	&	2.7	&	2.2	&	2.7	&	66	&	67	&	81	&	71	&	8	&	3	&	\nodata	\\
33523	&	B2V	&	8.83	&	-0.66	&	21320	&	4.5	&	4.8	&	4.8	&	166	&	196	&	166	&	176	&	17	&	3	&	A	\\
33554	&	B2/B3V	&	8.54	&	-0.55	&	17820	&	4.3	&	4.4	&	4.6	&	158	&	173	&	156	&	162	&	9	&	3	&	\nodata	\\
33575	&	B2V	&	5.58	&	-0.58	&	18680	&	2.6	&	2.0	&	2.0	&	59	&	54	&	56	&	57	&	3	&	3	&	\nodata	\\
33591	&	B3V	&	6.39	&	-0.5	&	16540	&	1.3	&	1.2	&	0.8	&	10	&	12	&	\nodata	&	11	&	1	&	2	&	\nodata	\\
33611	&	B2V	&	7.17	&	-0.66	&	21320	&	1.7	&	1.1	&	1.0	&	6	&	7	&	\nodata	&	6	&	1	&	2	&	A	\\
33635	&	B3V	&	8.71	&	-0.5	&	16540	&	1.9	&	1.5	&	1.3	&	36	&	30	&	27	&	31	&	5	&	3	&	\nodata	\\
33663	&	B3V	&	8.80	&	-0.58	&	18680	&	2.2	&	1.9	&	1.9	&	38	&	52	&	54	&	48	&	9	&	3	&	\nodata	\\
33703	&	B3V	&	6.30	&	-0.47	&	15840	&	3.4	&	3.2	&	3.8	&	113	&	117	&	124	&	118	&	5	&	3	&	\nodata	\\
33708	&	B3V	&	8.30	&	-0.5	&	16540	&	6.1	&	6.7	&	6.6	&	256	&	275	&	254	&	262	&	11	&	3	&	\nodata	\\
33769	&	B2/B3Ve	&	7.82	&	-0.6	&	19290	&	6.8	&	\nodata	&	7.0	&	294	&	\nodata	&	272	&	283	&	16	&	2	&	A	\\
33814	&	B3V	&	8.09	&	-0.65	&	20960	&	2.8	&	2.2	&	2.4	&	66	&	67	&	71	&	68	&	2	&	3	&	A	\\
33836	&	O9V	&	7.68	&	-0.89	&	33110	&	\nodata	&	6.2	&	7.6	&	\nodata	&	265	&	318	&	291	&	37	&	2	&	\nodata	\\
33846	&	B3Ve	&	6.94	&	-0.58	&	18680	&	5.0	&	5.5	&	5.3	&	195	&	225	&	191	&	204	&	18	&	3	&	A	\\
34041	&	B2/B3V	&	6.85	&	-0.6	&	19290	&	3.0	&	2.4	&	3.1	&	79	&	76	&	90	&	82	&	7	&	3	&	A	\\
34133	&	B0V	&	7.34	&	-0.82	&	28690	&	1.2	&	1.0	&	1.1	&	\nodata	&	9	&	5	&	7	&	3	&	2	&	C	\\
34213	&	B3V	&	9.38	&	-0.63	&	20260	&	1.0	&	1.1	&	1.1	&	2	&	8	&	0	&	3	&	4	&	3	&	\nodata	\\
34248	&	B3V	&	6.33	&	-0.52	&	17030	&	1.5	&	1.1	&	0.8	&	15	&	\nodata	&	\nodata	&	15	&	\nodata	&	1	&	\nodata	\\
34325	&	B1V	&	7.66	&	-0.78	&	26530	&	1.1	&	1.0	&	1.0	&	2	&	\nodata	&	3	&	3	&	0	&	2	&	\nodata	\\
34339	&	B3V	&	5.78	&	-0.56	&	18100	&	4.3	&	4.3	&	4.7	&	158	&	169	&	162	&	163	&	5	&	3	&	\nodata	\\
34350	&	B4V	&	6.47	&	-0.45	&	15410	&	2.0	&	1.6	&	1.3	&	\nodata	&	33	&	35	&	34	&	1	&	2	&	\nodata	\\
34395	&	B2V	&	8.72	&	-0.69	&	22470	&	2.8	&	2.1	&	2.4	&	72	&	67	&	69	&	69	&	3	&	3	&	A	\\
34478	&	B3V	&	9.00	&	-0.6	&	19290	&	2.1	&	1.8	&	1.9	&	\nodata	&	44	&	49	&	47	&	3	&	2	&	\nodata	\\
34489	&	B2V	&	6.65	&	-0.65	&	20960	&	3.3	&	2.9	&	3.1	&	96	&	104	&	90	&	96	&	7	&	3	&	A	\\
34499	&	B2V	&	8.75	&	-0.66	&	21320	&	1.9	&	1.3	&	1.1	&	14	&	16	&	\nodata	&	15	&	2	&	2	&	A	\\
34519	&	B2V	&	7.47	&	-0.72	&	23710	&	3.8	&	3.9	&	4.1	&	135	&	154	&	139	&	143	&	10	&	3	&	\nodata	\\
34562	&	B2V	&	8.27	&	-0.55	&	17820	&	1.9	&	1.5	&	1.3	&	25	&	30	&	25	&	27	&	3	&	3	&	\nodata	\\
34601	&	B2V	&	7.31	&	-0.62	&	19930	&	4.8	&	4.4	&	4.7	&	186	&	178	&	160	&	175	&	13	&	3	&	A	\\
34669	&	B4V	&	7.42	&	-0.48	&	16070	&	1.9	&	1.6	&	1.4	&	37	&	33	&	37	&	35	&	2	&	3	&	C	\\
34878	&	B3V	&	8.99	&	-0.59	&	18980	&	3.3	&	2.6	&	2.9	&	95	&	89	&	86	&	90	&	4	&	3	&	\nodata	\\
34894	&	B3V	&	8.49	&	-0.6	&	19290	&	3.5	&	2.9	&	3.1	&	106	&	101	&	93	&	100	&	7	&	3	&	\nodata	\\
34983	&	B3V	&	7.92	&	-0.58	&	18680	&	2.0	&	1.6	&	1.3	&	26	&	33	&	17	&	25	&	8	&	3	&	\nodata	\\
35083	&	B3V	&	5.34	&	-0.52	&	17030	&	1.4	&	1.5	&	1.2	&	\nodata	&	27	&	19	&	23	&	5	&	2	&	C	\\
35208	&	B3V	&	7.55	&	-0.6	&	19290	&	4.2	&	4.1	&	4.5	&	150	&	164	&	147	&	153	&	9	&	3	&	\nodata	\\
35226	&	B2IV-V	&	5.01	&	-0.57	&	18380	&	2.7	&	2.3	&	2.6	&	68	&	70	&	77	&	72	&	5	&	3	&	C	\\
35413	&	B3V	&	8.60	&	-0.49	&	16300	&	1.7	&	1.6	&	1.3	&	26	&	35	&	31	&	31	&	4	&	3	&	A	\\
35461	&	B2V	&	6.80	&	-0.61	&	19600	&	3.3	&	2.8	&	3.3	&	93	&	99	&	96	&	96	&	3	&	3	&	C	\\
35493	&	O9V	&	8.79	&	-0.88	&	32420	&	2.0	&	1.6	&	1.4	&	46	&	45	&	30	&	41	&	9	&	3	&	C	\\
35609	&	B3V	&	6.57	&	-0.52	&	17030	&	5.0	&	4.6	&	5.2	&	193	&	184	&	184	&	187	&	5	&	3	&	\nodata	\\
35683	&	B4V	&	8.40	&	-0.51	&	16780	&	3.3	&	2.9	&	3.4	&	108	&	103	&	109	&	107	&	3	&	3	&	\nodata	\\
35707	&	O9V	&	6.40	&	-0.9	&	33820	&	\nodata	&	\nodata	&	\nodata	&	\nodata	&	\nodata	&	\nodata	&	\nodata	&	\nodata	&	0	&	\nodata	\\
36096	&	B2V	&	8.85	&	-0.6	&	19290	&	4.4	&	4.2	&	4.8	&	161	&	166	&	163	&	163	&	3	&	3	&	\nodata	\\
36143	&	B2IV-V	&	5.88	&	-0.57	&	18380	&	4.3	&	4.2	&	4.7	&	156	&	166	&	159	&	160	&	5	&	3	&	\nodata	\\
36320	&	B3V	&	6.62	&	-0.49	&	16300	&	3.9	&	3.8	&	4.0	&	138	&	145	&	132	&	138	&	7	&	3	&	\nodata	\\
36582	&	B4V	&	6.65	&	-0.48	&	16070	&	2.0	&	1.8	&	1.4	&	44	&	44	&	35	&	41	&	5	&	3	&	\nodata	\\
36615	&	B1V	&	8.82	&	-0.78	&	26530	&	1.4	&	0.9	&	0.9	&	2	&	2	&	\nodata	&	2	&	0	&	2	&	\nodata	\\
36944	&	B3V	&	8.62	&	-0.6	&	19290	&	3.6	&	2.7	&	3.0	&	110	&	96	&	89	&	98	&	11	&	3	&	A	\\
36972	&	B3V	&	8.73	&	-0.6	&	19290	&	4.5	&	4.4	&	4.6	&	168	&	176	&	155	&	166	&	11	&	3	&	A	\\
37034	&	B3IV	&	7.20	&	-0.6	&	19290	&	2.6	&	2.0	&	1.9	&	55	&	53	&	51	&	53	&	2	&	3	&	\nodata	\\
37044	&	B4V	&	8.45	&	-0.54	&	17550	&	6.0	&	6.9	&	6.6	&	249	&	285	&	252	&	262	&	20	&	3	&	\nodata	\\
37245	&	B4/B5V	&	8.50	&	-0.55	&	17820	&	2.7	&	2.2	&	2.2	&	68	&	65	&	68	&	67	&	2	&	3	&	R	\\
37297	&	B2.5V	&	4.83	&	-0.51	&	16780	&	2.9	&	2.4	&	2.7	&	84	&	77	&	85	&	82	&	5	&	3	&	C	\\
37304	&	B6V	&	6.77	&	-0.44	&	15200	&	3.7	&	3.8	&	3.8	&	129	&	143	&	127	&	133	&	9	&	3	&	\nodata	\\
37439	&	B3V	&	8.01	&	-0.49	&	16300	&	2.6	&	2.2	&	2.3	&	72	&	66	&	73	&	70	&	4	&	3	&	\nodata	\\
37597	&	B3V	&	7.64	&	-0.54	&	17550	&	1.1	&	1.4	&	1.0	&	\nodata	&	22	&	\nodata	&	22	&	\nodata	&	1	&	\nodata	\\
37803	&	B3IV	&	6.53	&	-0.59	&	18980	&	1.8	&	2.0	&	1.1	&	9	&	\nodata	&	8	&	9	&	1	&	2	&	C	\\
37995	&	B1/B2V	&	5.87	&	-0.68	&	22070	&	2.4	&	1.8	&	1.9	&	48	&	51	&	50	&	50	&	2	&	3	&	R	\\
38028	&	B3V	&	6.98	&	-0.63	&	20260	&	3.0	&	2.6	&	3.3	&	77	&	87	&	96	&	86	&	9	&	3	&	\nodata	\\
38457	&	B3V	&	8.26	&	-0.47	&	15840	&	3.5	&	2.9	&	2.9	&	118	&	104	&	92	&	105	&	13	&	3	&	\nodata	\\
38593	&	B2V	&	5.46	&	-0.6	&	19290	&	2.7	&	2.0	&	2.3	&	61	&	58	&	66	&	62	&	4	&	3	&	\nodata	\\
38716	&	B0V	&	10.68	&	-0.85	&	30480	&	1.7	&	1.2	&	1.3	&	19	&	26	&	22	&	22	&	3	&	3	&	C	\\
38727	&	B3V	&	7.99	&	-0.6	&	19290	&	3.7	&	3.3	&	3.7	&	121	&	122	&	111	&	118	&	6	&	3	&	\nodata	\\
38795	&	B2V	&	6.77	&	-0.64	&	20600	&	2.4	&	1.9	&	1.9	&	45	&	53	&	47	&	48	&	4	&	3	&	A	\\
38858	&	B3V	&	7.28	&	-0.53	&	17290	&	3.8	&	3.5	&	3.5	&	129	&	131	&	109	&	123	&	12	&	3	&	\nodata	\\
38896	&	B3Ve	&	7.24	&	-0.55	&	17820	&	2.4	&	2.1	&	2.0	&	54	&	61	&	60	&	59	&	4	&	3	&	\nodata	\\
38942	&	B4V	&	7.47	&	-0.48	&	16070	&	2.9	&	2.7	&	2.9	&	90	&	94	&	93	&	92	&	3	&	3	&	\nodata	\\
39014	&	B4V	&	5.96	&	-0.52	&	17030	&	4.1	&	3.9	&	4.2	&	144	&	152	&	140	&	145	&	6	&	3	&	\nodata	\\
39063	&	B2/B3V	&	7.41	&	-0.51	&	16780	&	1.8	&	1.6	&	1.4	&	27	&	31	&	35	&	31	&	4	&	3	&	\nodata	\\
39138	&	B3V	&	4.80	&	-0.5	&	16540	&	1.7	&	1.3	&	0.9	&	23	&	15	&	\nodata	&	19	&	6	&	2	&	\nodata	\\
39238	&	B4V	&	6.27	&	-0.52	&	17030	&	3.0	&	2.7	&	3.1	&	91	&	95	&	97	&	95	&	3	&	3	&	\nodata	\\
39446	&	B4V	&	8.55	&	-0.54	&	17550	&	2.9	&	2.9	&	2.9	&	82	&	105	&	90	&	92	&	11	&	3	&	\nodata	\\
39483	&	B3V	&	7.61	&	-0.44	&	15200	&	3.3	&	2.4	&	3.2	&	114	&	78	&	103	&	99	&	18	&	3	&	\nodata	\\
39540	&	B1V	&	7.81	&	-0.78	&	26530	&	1.3	&	1.0	&	0.9	&	\nodata	&	4	&	4	&	4	&	1	&	2	&	\nodata	\\
39613	&	B3V	&	7.05	&	-0.56	&	18100	&	4.5	&	4.3	&	5.0	&	168	&	170	&	177	&	172	&	4	&	3	&	\nodata	\\
39774	&	B3V	&	8.14	&	-0.62	&	19930	&	2.0	&	1.4	&	1.2	&	19	&	25	&	\nodata	&	22	&	4	&	2	&	\nodata	\\
39782	&	B3V	&	8.64	&	-0.6	&	19290	&	2.3	&	2.0	&	2.3	&	\nodata	&	55	&	67	&	61	&	8	&	2	&	\nodata	\\
39880	&	B4V	&	8.14	&	-0.52	&	17030	&	2.5	&	2.2	&	2.2	&	61	&	64	&	68	&	64	&	3	&	3	&	\nodata	\\
40077	&	B3V	&	5.81	&	-0.51	&	16780	&	2.6	&	2.2	&	2.7	&	68	&	68	&	85	&	74	&	10	&	3	&	\nodata	\\
40265	&	B3V	&	7.53	&	-0.64	&	20600	&	3.1	&	2.6	&	2.7	&	83	&	90	&	78	&	84	&	6	&	3	&	\nodata	\\
40268	&	B2/B3V	&	7.34	&	-0.54	&	17550	&	\nodata	&	7.1	&	\nodata	&	\nodata	&	293	&	\nodata	&	293	&	\nodata	&	1	&	C	\\
40299	&	B2V	&	6.48	&	-0.58	&	18680	&	3.9	&	3.8	&	4.1	&	132	&	148	&	133	&	137	&	9	&	3	&	\nodata	\\
40366	&	B3V	&	6.57	&	-0.48	&	16070	&	2.9	&	2.7	&	2.9	&	92	&	92	&	92	&	92	&	0	&	3	&	\nodata	\\
40629	&	B3V	&	8.70	&	-0.6	&	19290	&	3.9	&	3.9	&	4.1	&	127	&	153	&	132	&	137	&	14	&	3	&	\nodata	\\
41323	&	B3V	&	5.42	&	-0.51	&	16780	&	2.0	&	1.7	&	1.5	&	38	&	39	&	36	&	38	&	1	&	3	&	\nodata	\\
41463	&	B2V	&	6.66	&	-0.54	&	17550	&	3.0	&	2.5	&	2.6	&	88	&	84	&	80	&	84	&	4	&	3	&	R	\\
41640	&	B4V	&	6.68	&	-0.46	&	15630	&	1.8	&	1.6	&	1.3	&	35	&	32	&	31	&	33	&	2	&	3	&	\nodata	\\
41680	&	B4V	&	9.08	&	-0.48	&	16070	&	2.2	&	1.9	&	1.9	&	51	&	51	&	61	&	54	&	6	&	3	&	\nodata	\\
41737	&	B4IV	&	6.29	&	-0.49	&	16300	&	3.6	&	3.8	&	4.1	&	124	&	144	&	135	&	134	&	10	&	3	&	\nodata	\\
41823	&	B3V	&	6.98	&	-0.6	&	19290	&	3.2	&	3.5	&	3.5	&	92	&	132	&	104	&	109	&	21	&	3	&	R	\\
41862	&	B3Ve	&	9.49	&	-0.56	&	18100	&	6.4	&	6.3	&	6.0	&	273	&	261	&	227	&	254	&	24	&	3	&	\nodata	\\
41970	&	B2.5V	&	6.46	&	-0.54	&	17550	&	2.7	&	2.1	&	2.5	&	73	&	63	&	76	&	71	&	7	&	3	&	\nodata	\\
42038	&	B3V	&	6.79	&	-0.52	&	17030	&	2.1	&	1.9	&	1.9	&	42	&	52	&	57	&	51	&	8	&	3	&	R	\\
42357	&	B4V	&	8.69	&	-0.54	&	17550	&	1.2	&	1.1	&	0.8	&	4	&	6	&	\nodata	&	5	&	1	&	2	&	C	\\
42595	&	B2IV	&	6.93	&	-0.71	&	23280	&	4.7	&	5.0	&	5.2	&	183	&	205	&	189	&	192	&	11	&	3	&	\nodata	\\
42653	&	B3V	&	6.69	&	-0.49	&	16300	&	4.8	&	4.3	&	4.7	&	183	&	167	&	162	&	171	&	11	&	3	&	A	\\
42698	&	B3V	&	8.28	&	-0.54	&	17550	&	3.9	&	3.8	&	4.0	&	134	&	146	&	128	&	136	&	9	&	3	&	A	\\
42868	&	B3IV	&	7.21	&	-0.53	&	17290	&	3.6	&	3.3	&	3.4	&	120	&	124	&	108	&	117	&	8	&	3	&	\nodata	\\
42908	&	B1.5V	&	7.32	&	-0.76	&	25540	&	2.5	&	3.0	&	3.7	&	\nodata	&	115	&	120	&	118	&	3	&	2	&	C	\\
43114	&	B4V	&	6.32	&	-0.47	&	15840	&	4.0	&	3.6	&	3.5	&	145	&	134	&	112	&	130	&	17	&	3	&	R	\\
43285	&	B3V	&	7.30	&	-0.49	&	16300	&	3.0	&	2.9	&	3.1	&	94	&	101	&	98	&	98	&	3	&	3	&	A	\\
43464	&	B1.5V	&	7.80	&	-0.73	&	24150	&	3.4	&	3.0	&	3.2	&	110	&	114	&	98	&	107	&	9	&	3	&	C	\\
43473	&	B3V	&	7.69	&	-0.6	&	19290	&	3.6	&	3.4	&	3.6	&	112	&	127	&	107	&	115	&	10	&	3	&	\nodata	\\
43520	&	B3V	&	6.36	&	-0.49	&	16300	&	3.4	&	3.5	&	3.4	&	113	&	130	&	108	&	117	&	11	&	3	&	A	\\
43699	&	B4V	&	8.20	&	-0.48	&	16070	&	5.4	&	4.4	&	5.4	&	214	&	176	&	195	&	195	&	19	&	3	&	\nodata	\\
44251	&	B6Vn	&	7.26	&	-0.48	&	16070	&	5.2	&	4.7	&	5.2	&	207	&	189	&	186	&	194	&	11	&	3	&	R	\\
44509	&	B3IV/V	&	6.43	&	-0.53	&	17290	&	1.9	&	1.5	&	1.4	&	26	&	29	&	32	&	29	&	3	&	3	&	\nodata	\\
44996	&	B4V	&	6.82	&	-0.46	&	15630	&	1.9	&	1.9	&	1.5	&	42	&	49	&	44	&	45	&	4	&	3	&	A	\\
45044	&	B3/B4V	&	7.46	&	-0.58	&	18680	&	3.9	&	3.9	&	4.1	&	133	&	152	&	130	&	138	&	12	&	3	&	\nodata	\\
45094	&	B4V	&	8.41	&	-0.53	&	17290	&	2.5	&	2.5	&	3.0	&	\nodata	&	82	&	94	&	88	&	8	&	3	&	\nodata	\\
46224	&	B4V	&	8.11	&	-0.51	&	16780	&	3.0	&	2.7	&	3.2	&	89	&	94	&	102	&	95	&	7	&	3	&	R	\\
46760	&	B2V	&	7.72	&	-0.72	&	23710	&	1.8	&	1.3	&	1.2	&	17	&	23	&	\nodata	&	20	&	4	&	2	&	R	\\
47137	&	B3V	&	8.11	&	-0.47	&	15840	&	4.5	&	4.7	&	4.3	&	168	&	190	&	143	&	167	&	24	&	3	&	\nodata	\\
48128	&	B2V	&	8.64	&	-0.63	&	20260	&	3.2	&	3.0	&	3.5	&	88	&	109	&	103	&	100	&	11	&	3	&	\nodata	\\
48469	&	B1IV	&	6.49	&	-0.84	&	29860	&	2.2	&	1.8	&	1.7	&	58	&	61	&	53	&	57	&	4	&	3	&	R	\\
48782	&	B3V	&	6.18	&	-0.53	&	17290	&	2.9	&	2.7	&	3.0	&	84	&	91	&	92	&	89	&	5	&	3	&	\nodata	\\
48835	&	B3V	&	6.05	&	-0.59	&	18980	&	4.1	&	3.8	&	4.4	&	141	&	149	&	147	&	146	&	4	&	3	&	R	\\
49201	&	B2V	&	7.70	&	-0.72	&	23710	&	3.7	&	3.7	&	3.6	&	126	&	145	&	113	&	128	&	16	&	3	&	\nodata	\\
50067	&	B2V	&	6.41	&	-0.6	&	19290	&	1.5	&	1.1	&	0.8	&	5	&	3	&	\nodata	&	4	&	1	&	2	&	\nodata	\\
50126	&	B2/B3IV	&	7.68	&	-0.6	&	19290	&	1.6	&	1.2	&	1.1	&	4	&	\nodata	&	6	&	6	&	1	&	2	&	\nodata	\\
50135	&	B4V	&	7.48	&	-0.52	&	17030	&	2.4	&	2.0	&	1.9	&	61	&	58	&	56	&	58	&	2	&	3	&	C	\\
52202	&	B4V	&	8.00	&	-0.5	&	16540	&	4.4	&	4.2	&	4.4	&	164	&	163	&	146	&	158	&	10	&	3	&	\nodata	\\
52444	&	O9V	&	8.36	&	-0.91	&	34550	&	\nodata	&	\nodata	&	\nodata	&	\nodata	&	\nodata	&	\nodata	&	\nodata	&	\nodata	&	0	&	\nodata	\\
52766	&	B4V	&	7.50	&	-0.5	&	16540	&	1.9	&	1.8	&	1.8	&	\nodata	&	46	&	53	&	50	&	5	&	2	&	\nodata	\\
52868	&	B3V	&	7.81	&	-0.58	&	18680	&	4.1	&	3.8	&	4.2	&	140	&	147	&	136	&	141	&	6	&	3	&	\nodata	\\
52977	&	B3V	&	7.20	&	-0.5	&	16540	&	4.3	&	4.4	&	4.7	&	157	&	173	&	162	&	164	&	8	&	3	&	\nodata	\\
53018	&	B4V	&	7.80	&	-0.5	&	16540	&	1.4	&	1.1	&	0.7	&	13	&	6	&	\nodata	&	10	&	5	&	3	&	\nodata	\\
53057	&	B4V	&	7.47	&	-0.5	&	16540	&	1.5	&	1.4	&	1.0	&	16	&	\nodata	&	13	&	14	&	2	&	2	&	\nodata	\\
53686	&	O8/O9	&	8.57	&	-0.91	&	34550	&	\nodata	&	\nodata	&	\nodata	&	\nodata	&	\nodata	&	\nodata	&	\nodata	&	\nodata	&	0	&	\nodata	\\
54175	&	O9V	&	7.63	&	-0.9	&	33820	&	\nodata	&	\nodata	&	\nodata	&	\nodata	&	\nodata	&	\nodata	&	\nodata	&	\nodata	&	0	&	\nodata	\\
54930	&	B0.5IVn	&	8.81	&	-0.89	&	33110	&	\nodata	&	7.6	&	\nodata	&	\nodata	&	326	&	\nodata	&	326	&	\nodata	&	1	&	\nodata	\\
55051	&	B1V	&	7.39	&	-0.75	&	25060	&	3.5	&	3.4	&	3.7	&	120	&	135	&	120	&	125	&	9	&	3	&	\nodata	\\
55078	&	O9V	&	\nodata	&	-0.72	&	23710	&	2.3	&	\nodata	&	1.6	&	48	&	\nodata	&	36	&	42	&	8	&	2	&	\nodata	\\
55938	&	B3V	&	7.65	&	-0.55	&	17820	&	2.6	&	2.4	&	2.8	&	63	&	77	&	85	&	75	&	11	&	3	&	\nodata	\\
55977	&	B3V	&	10.60	&	-0.62	&	19930	&	6.8	&	7.3	&	\nodata	&	293	&	303	&	\nodata	&	298	&	7	&	2	&	\nodata	\\
57848	&	B4V	&	7.19	&	-0.48	&	16070	&	1.4	&	1.4	&	1.1	&	16	&	22	&	21	&	20	&	4	&	3	&	\nodata	\\
58128	&	B2Ve	&	7.98	&	-0.57	&	18380	&	5.6	&	5.4	&	5.4	&	228	&	222	&	198	&	216	&	16	&	3	&	\nodata	\\
58326	&	B3V	&	5.55	&	-0.56	&	18100	&	3.8	&	3.7	&	4.4	&	126	&	144	&	145	&	138	&	10	&	3	&	\nodata	\\
59288	&	O9V	&	8.18	&	-0.91	&	34550	&	\nodata	&	\nodata	&	\nodata	&	\nodata	&	\nodata	&	\nodata	&	\nodata	&	\nodata	&	0	&	\nodata	\\
59830	&	B3V	&	8.35	&	-0.57	&	18380	&	2.5	&	2.2	&	2.2	&	58	&	67	&	66	&	64	&	5	&	3	&	C	\\
60429	&	B0—O9V	&	8.71	&	-0.84	&	29860	&	2.8	&	2.5	&	2.8	&	94	&	95	&	91	&	94	&	2	&	3	&	\nodata	\\
62327	&	B3V	&	4.61	&	-0.51	&	16780	&	1.7	&	1.5	&	1.2	&	22	&	26	&	20	&	22	&	3	&	3	&	A	\\
63517	&	B2V	&	9.88	&	-0.72	&	23710	&	3.9	&	3.6	&	4.1	&	138	&	144	&	136	&	139	&	4	&	3	&	\nodata	\\
64716	&	B1/B2V	&	8.81	&	-0.75	&	25060	&	5.1	&	5.5	&	6.0	&	214	&	229	&	234	&	225	&	11	&	3	&	A	\\
67279	&	B2V	&	7.82	&	-0.66	&	21320	&	2.8	&	2.5	&	3.2	&	69	&	84	&	94	&	82	&	13	&	3	&	\nodata	\\
67969	&	B3V	&	8.25	&	-0.6	&	19290	&	3.5	&	3.7	&	3.7	&	106	&	140	&	113	&	120	&	18	&	3	&	\nodata	\\
68124	&	B2/B3IV	&	6.93	&	-0.58	&	18680	&	3.9	&	3.9	&	4.4	&	130	&	151	&	143	&	142	&	11	&	3	&	\nodata	\\
68829	&	B3V	&	8.74	&	-0.5	&	16540	&	0.9	&	0.7	&	0.7	&	0	&	\nodata	&	\nodata	&	0	&	\nodata	&	1	&	\nodata	\\
68862	&	B2V	&	4.34	&	-0.63	&	20260	&	2.0	&	1.4	&	1.2	&	17	&	23	&	9	&	16	&	7	&	3	&	A	\\
68992	&	B5IV/V	&	9.42	&	-0.51	&	16780	&	2.7	&	2.6	&	2.6	&	76	&	87	&	81	&	81	&	6	&	3	&	\nodata	\\
69617	&	B2V	&	9.55	&	-0.75	&	25060	&	2.8	&	2.7	&	2.7	&	80	&	100	&	82	&	87	&	11	&	3	&	\nodata	\\
69640	&	B1V:	&	9.97	&	-0.78	&	26530	&	2.9	&	2.7	&	3.3	&	88	&	102	&	105	&	98	&	9	&	3	&	\nodata	\\
70477	&	B4V	&	8.60	&	-0.54	&	17550	&	3.9	&	3.0	&	3.9	&	137	&	108	&	124	&	123	&	14	&	3	&	\nodata	\\
70506	&	B1V	&	8.98	&	-0.77	&	26030	&	1.8	&	1.3	&	1.1	&	20	&	22	&	\nodata	&	21	&	1	&	2	&	\nodata	\\
71666	&	B3IV	&	9.29	&	-0.53	&	17290	&	3.0	&	2.5	&	3.0	&	88	&	81	&	92	&	87	&	5	&	3	&	\nodata	\\
71763	&	B4V	&	10.11	&	-0.54	&	17550	&	3.7	&	3.6	&	3.8	&	125	&	137	&	119	&	127	&	9	&	3	&	\nodata	\\
73494	&	B4V	&	6.99	&	-0.54	&	17550	&	1.2	&	1.0	&	0.7	&	3	&	3	&	\nodata	&	3	&	1	&	2	&	\nodata	\\
73624	&	B3V	&	5.43	&	-0.54	&	17550	&	1.6	&	1.3	&	0.8	&	15	&	19	&	\nodata	&	17	&	3	&	2	&	\nodata	\\
73881	&	B2/B3V	&	6.47	&	-0.57	&	18380	&	3.9	&	3.9	&	4.2	&	132	&	152	&	135	&	140	&	10	&	3	&	\nodata	\\
74110	&	B3V	&	6.82	&	-0.53	&	17290	&	1.8	&	1.4	&	1.1	&	22	&	24	&	14	&	20	&	5	&	3	&	\nodata	\\
74784	&	B3V	&	8.00	&	-0.54	&	22870	&	5.5	&	5.7	&	6.0	&	227	&	238	&	231	&	232	&	6	&	3	&	C	\\
75091	&	B3V	&	6.32	&	-0.54	&	17550	&	3.2	&	3.1	&	3.4	&	97	&	111	&	106	&	105	&	7	&	3	&	\nodata	\\
75304	&	B4V	&	4.52	&	-0.51	&	16780	&	3.9	&	\nodata	&	4.3	&	137	&	\nodata	&	145	&	141	&	6	&	2	&	A	\\
75959	&	B3V	&	6.38	&	-0.6	&	19290	&	1.3	&	1.1	&	0.9	&	\nodata	&	8	&	\nodata	&	8	&	\nodata	&	1	&	R	\\
76126	&	B3V	&	5.49	&	-0.66	&	21320	&	4.8	&	\nodata	&	5.8	&	189	&	\nodata	&	216	&	203	&	19	&	2	&	\nodata	\\
77859	&	B2V	&	5.40	&	-0.56	&	18100	&	5.8	&	5.8	&	6.5	&	238	&	238	&	248	&	241	&	6	&	3	&	A	\\
78933	&	B1V	&	3.95	&	-0.78	&	26530	&	2.9	&	2.7	&	3.1	&	88	&	102	&	96	&	95	&	7	&	3	&	A	\\
80461	&	B3/B4V	&	6.78	&	-0.57	&	18380	&	3.4	&	3.1	&	3.7	&	102	&	114	&	115	&	111	&	7	&	3	&	A	\\
80815	&	B3V	&	4.78	&	-0.61	&	19600	&	4.5	&	4.4	&	4.8	&	168	&	174	&	165	&	169	&	4	&	3	&	A	\\
81214	&	B3V	&	8.35	&	-0.64	&	20600	&	2.9	&	2.4	&	2.8	&	74	&	78	&	81	&	78	&	3	&	3	&	\nodata	\\
81266	&	B0.2V	&	2.81	&	-0.84	&	29860	&	1.1	&	0.7	&	0.8	&	4	&	\nodata	&	1	&	3	&	2	&	2	&	A	\\
81972	&	B3V	&	5.64	&	-0.55	&	17820	&	4.5	&	4.4	&	4.7	&	165	&	177	&	159	&	167	&	9	&	3	&	A	\\
82034	&	B3V	&	7.53	&	-0.6	&	19290	&	2.0	&	2.0	&	1.9	&	\nodata	&	57	&	51	&	50	&	9	&	2	&	\nodata	\\
82254	&	B4V	&	6.81	&	-0.44	&	15200	&	3.5	&	4.0	&	3.9	&	123	&	153	&	129	&	135	&	16	&	3	&	\nodata	\\
83506	&	B2V	&	8.83	&	-0.61	&	19600	&	4.9	&	5.1	&	5.3	&	190	&	208	&	190	&	196	&	10	&	3	&	\nodata	\\
83509	&	B2V	&	8.32	&	-0.65	&	20960	&	4.1	&	3.9	&	4.6	&	141	&	151	&	156	&	149	&	8	&	3	&	\nodata	\\
83635	&	B1V	&	5.64	&	-0.78	&	26530	&	3.2	&	3.1	&	3.7	&	108	&	122	&	121	&	117	&	8	&	3	&	R	\\
83861	&	B2V	&	8.73	&	-0.78	&	26530	&	3.3	&	3.2	&	3.7	&	111	&	124	&	123	&	119	&	7	&	3	&	\nodata	\\
84435	&	B3V	&	8.68	&	-0.53	&	17290	&	2.7	&	2.3	&	2.6	&	75	&	69	&	81	&	75	&	6	&	3	&	\nodata	\\
85720	&	B0V	&	10.31	&	-0.88	&	32420	&	1.5	&	1.1	&	1.2	&	\nodata	&	15	&	15	&	15	&	0	&	3	&	\nodata	\\
86349	&	B1V	&	9.04	&	-0.74	&	24600	&	3.3	&	3.1	&	3.7	&	106	&	121	&	118	&	115	&	8	&	3	&	\nodata	\\
86508	&	B4V	&	7.24	&	-0.53	&	17290	&	3.3	&	3.3	&	3.8	&	102	&	121	&	122	&	115	&	11	&	3	&	C	\\
86951	&	B3V	&	8.45	&	-0.57	&	18380	&	3.6	&	3.5	&	4.1	&	118	&	134	&	131	&	127	&	9	&	3	&	\nodata	\\
86954	&	B4	&	6.84	&	-0.55	&	17820	&	2.2	&	1.7	&	1.6	&	42	&	40	&	41	&	41	&	1	&	3	&	C	\\
87218	&	B3V	&	8.59	&	-0.49	&	16300	&	4.1	&	4.1	&	4.6	&	145	&	159	&	158	&	154	&	8	&	3	&	\nodata	\\
87508	&	B2/B3V	&	8.59	&	-0.56	&	18100	&	3.6	&	3.5	&	3.9	&	114	&	130	&	122	&	122	&	8	&	3	&	\nodata	\\
88201	&	B3V	&	8.07	&	-0.6	&	19290	&	3.7	&	3.6	&	3.9	&	118	&	139	&	120	&	126	&	12	&	3	&	R	\\
88857	&	B3V	&	7.67	&	-0.53	&	17290	&	4.2	&	4.4	&	3.9	&	152	&	173	&	125	&	150	&	24	&	3	&	\nodata	\\
89551	&	B2V	&	7.57	&	-0.72	&	23710	&	1.9	&	1.3	&	1.3	&	21	&	20	&	15	&	19	&	3	&	3	&	\nodata	\\
89902	&	B2V	&	6.97	&	-0.72	&	23710	&	1.9	&	1.4	&	1.0	&	17	&	27	&	\nodata	&	22	&	7	&	2	&	R	\\
90676	&	B2V	&	6.69	&	-0.58	&	18680	&	1.4	&	1.0	&	0.8	&	4	&	\nodata	&	2	&	3	&	1	&	2	&	\nodata	\\
91038	&	B1V	&	8.70	&	-0.78	&	26530	&	1.7	&	1.2	&	1.1	&	12	&	18	&	\nodata	&	15	&	4	&	2	&	\nodata	\\
91918	&	B2.5V	&	4.84	&	-0.58	&	18680	&	1.9	&	1.4	&	1.2	&	22	&	22	&	\nodata	&	22	&	0	&	2	&	\nodata	\\
92393	&	B3V	&	6.67	&	-0.59	&	18980	&	1.6	&	1.4	&	1.2	&	\nodata	&	21	&	12	&	16	&	6	&	2	&	\nodata	\\
92904	&	B2V	&	7.35	&	-0.66	&	21320	&	2.8	&	2.6	&	2.9	&	71	&	88	&	85	&	81	&	9	&	3	&	\nodata	\\
92957	&	B3V	&	8.03	&	-0.57	&	18380	&	4.3	&	4.5	&	4.8	&	155	&	180	&	167	&	167	&	12	&	3	&	\nodata	\\
93996	&	B2Ve	&	5.57	&	-0.61	&	19600	&	3.2	&	2.5	&	3.0	&	89	&	80	&	87	&	86	&	5	&	3	&	\nodata	\\
94385	&	B3V	&	5.36	&	-0.51	&	16780	&	3.8	&	3.7	&	4.0	&	129	&	140	&	131	&	133	&	6	&	3	&	R	\\
94513	&	B2V	&	10.19	&	-0.71	&	23280	&	2.0	&	1.6	&	1.4	&	26	&	\nodata	&	25	&	26	&	1	&	2	&	\nodata	\\
97162	&	B2V	&	\nodata	&	-0.64	&	20600	&	1.9	&	\nodata	&	1.5	&	10	&	\nodata	&	28	&	19	&	13	&	2	&	\nodata	\\
97680	&	B3V	&	10.19	&	-0.55	&	17820	&	2.8	&	2.5	&	3.0	&	77	&	82	&	91	&	83	&	7	&	3	&	R	\\
100170	&	B2V	&	8.83	&	-0.68	&	22070	&	2.3	&	1.8	&	1.7	&	40	&	49	&	43	&	44	&	5	&	3	&	\nodata	\\
101909	&	B3V	&	5.98	&	-0.59	&	18980	&	4.0	&	3.9	&	4.4	&	139	&	151	&	144	&	145	&	6	&	3	&	R	\\

\enddata

\tablecomments{The columns are: (1) HIPPARCOS identification; (2) spectral type; (3) apparent magnitude $V$ (4) value of the $Q$ parameter; (5) effective temperature $T_{eff}$; (6, 7, 8) full width at half maximum in \AA\ of the three He {\sc i} lines ($\lambda4026$, $\lambda4388$ and $\lambda4471$ \AA); (9, 10, 11) $v\sin i$ of the same He {\sc i} lines; (12) mean $v\sin i$, (13) standard deviation of the $v\sin i$ between the availbale measures; (14) the number of He {\sc i} lines used, and (15) the membership classification (`A' for associations, `C' for clusters and `R' for runaway stars). All the velocities are in km s$^{-1}$}
\tablecomments{Table \ref{tab:sample} is published in its entirety in the 
electronic edition of the {\it Astrophysical Journal}.  A portion is 
shown here for guidance regarding its form and content.}

\end{deluxetable}